\documentclass[a4paper,11pt]{article}
\usepackage{jheppub} 

\def\beq#1\eeq{\begin{align}#1\end{align}}

\title{ 4d $\mathcal{ N}=2$  SCFT and singularity theory Part I: Classification}

\author[b,c]{Dan Xie}
\author[a,b,c]{Shing-Tung Yau}

\affiliation[a]{Department of Mathematics, Harvard University, Cambridge, MA 02138, USA}
\affiliation[b]{Center of Mathematical Sciences and Applications, Harvard University, Cambridge, 02138, USA}
\affiliation[c]{Jefferson Physical Laboratory, Harvard University, Cambridge, MA 02138, USA}

\abstract{This is the first of a series of papers in which we systematically use singularity theory to study 
four dimensional $\mathcal{N}=2$ superconformal field theories. Our main focus in this paper is
to identify what kind of singularity is needed to define a SCFT. The constraint for a hypersurface singularity has 
been found by Sharpere and Vafa, and here the complete set of solutions are listed using a related mathematical  
result of Stephen S. T. Yau and Yu. We also study other type of singularities such as the 
complete intersection, quotient of hypersurface singularity by a finite group and non-isolated singularity. We finally conjecture that any three dimensional rational Gorenstein graded isolated singularity
should define a $\mathcal{N}=2$ SCFT. 
We explain how to extract various interesting physical quantities such as Seiberg-Witten geometry, central charges,  exact marginal deformations,  BPS quiver, RG flow trajectory, etc from the properties of 
singularity. }

\begin{document} 
\maketitle
\flushbottom

\section{Introduction}
The space of four dimensional
$\mathcal{N}=2$ superconformal field theory (SCFT) is becoming increasingly larger since the Seiberg-Witten (SW) solution was found \cite{Seiberg:1994rs,Seiberg:1994aj}.
 The two earlier examples studied in \cite{Seiberg:1994aj} are  $\text{SU}(2)$ gauge theory coupled with 4 fundamental flavors and $\text{SU}(2)$ gauge theory 
coupled with an adjoint hypermultiplet. It became immediately clear that one can find many new SCFTs with Lagrangian 
descriptions \cite{Argyres:1995wt, Witten:1997sc}, and it was also soon realized that more exotic SCFTs like Argyres-Douglas theories \cite{Argyres:1995jj, Argyres:1995xn, Eguchi:1996vu} exist. 
Quite recently, the space of $\mathcal{N}=2$ SCFT 
is greatly enlarged by the so-called class ${\cal S}$ construction in which one can engineered 4d $\mathcal{N}=2$ theory by putting 
6d $(2,0)$ theory on a punctured Riemann surface \cite{Gaiotto:2009we,Gaiotto:2009hg,Xie:2012hs,Wang:2015mra, Chacaltana:2012zy}.
Usually some of SCFTs in this class can be described by
non-abelian gauge groups coupled to various matter contents, such as free hypermultiplets, strongly coupled matter systems like $T_N$ theory and its cousins.

Given such rich set of theories found already, 
one might wonder if we can further enlarge the space of $\mathcal{N}=2$ SCFTs. One feature of above class is that 
SW solution of almost all the theories considered above is given by a curve fibered over the Coulomb branch moduli space, namely, 
the SW curve is put in the form $F(x,z, \lambda_i)=0$ where $\lambda_i$ are the parameter spaces of Coulomb branch including couplings, masses, and Coulomb branch moduli.
However, as pioneered and emphasized by Vafa and collaborators in a series of papers
\cite{Klemm:1996bj, Katz:1997eq, Gukov:1999ya,  Shapere:1999xr} back in 90s, the SW solution of general $\mathcal{N}=2$ theory should be  given by a three fold fibered over the 
moduli space: $F(z_0,z_1,z_2,z_3,\lambda_i)=0$, and only in special case the solution can be reduced to a curve fibration.
For example, the solution of quiver gauge theory with affine $E$ shape might only take a form of three fold fibration \cite{Katz:1997eq}. 
Recently, this approach has bee used in \cite{Cecotti:2010fi,Cecotti:2011gu, DelZotto:2011an,DelZotto:2015rca,Wang:2015mra} to find 
many new interesting theories. 

One could  greatly extend the space of $\mathcal{N}=2$ SCFTs by looking at all possible theories whose SW solution is given by three-fold fibration.
Now following the philosophy of geometric engineering \cite{Katz:1996fh,Shapere:1999xr}, one only 
need to start with a three-fold singularity, and
 the full SW solution is given by the deformation of the singularity \cite{Shapere:1999xr}. So the classification of 
 $\mathcal{N}=2$ theory is reduced to the classification of possible singularities, and this significantly simplifies the 
 task of classification. 

The main purpose of this paper is to try to use algebraic geometry of singularity theory to systematically classify 
possible $\mathcal{N}=2$ theory following the ideas in \cite{Shapere:1999xr}.
The constraints on isolated three-fold hypersurface singularities (IHS) defined by a polynomial $f(z_0,z_1, z_2, z_3)$ are already
described in \cite{Gukov:1999ya, Shapere:1999xr}: 
a key feature for $\mathcal{N}=2$ SCFT  is the existence of a $U(1)_R$ symmetry and geometrically this implies that
 the three-fold singularity should have  a $\mathbb{C}^*$ action:
\begin{equation}
f(\lambda^{q_i}z_i)=\lambda f(z_i),~~~q_i>0;
\end{equation}
Moreover to get a sensible SCFT, the weights of $\mathbb{C}^*$ action have to satisfy the following condition:
\begin{equation}
\sum q_i>1.
\label{cons}
\end{equation} 
Shapere and Vafa conjectured that that these are the necessary and  sufficient conditions for IHS to define a $\mathcal{N}=2$ SCFT \cite{Shapere:1999xr}. 
Therefore the classification of SCFTs arising from IHS  is reduced to the classification of  
IHS with $\mathbb{C}^*$ action whose weights satisfy \ref{cons}. 
Interestingly, such singularities 
have already been classified by Stephen S.T. Yau and Yu in a rather different mathematical context \cite{yau2003classification}, and we simply reorganize their results here.  

One of the most remarkable advantage of using singularity to define a SCFT is that the SW solution is automatically given 
by the mini-versal deformation of the singularity \cite{arnold2012singularities}.  Let's take $\phi_{\alpha}(z)$ as the monomial basis of the Jacobi algebra $\mathbb{C}[z_0,z_1,z_2, z_3]/({\partial f\over z_0},{\partial f\over z_1},{\partial f\over z_2},{\partial f\over z_3} )$, then 
the SW geometry is simply given by the following formula \cite{Shapere:1999xr}:
\begin{align}
&F(z,\lambda)=f(z)+\sum_{i=1}^\mu \lambda_\alpha \phi_\alpha(z), \nonumber\\
&\Omega={dz_0\wedge dz_1\wedge dz_2\wedge dz_3\over dF};
\end{align}
Here $\mu$ is the dimension of Jacobi algebra, and $\Omega$ is the SW differential. The scaling dimensions of $\lambda_\alpha$ can also be easily computed by requiring $\Omega$ having dimension one
as it gives the mass of BPS particles.
Once the SW solution
is given, we can study various physical quantities such as low energy effective action, central charges, BPS spectrum, RG flow, etc. It turns out that 
those properties are naturally related to the quantity studied in the  singularity theory \cite{arnold2012singularities, arnol?d2012singularities, looijenga1984isolated, dimca2012singularities}.

One can also consider other type of three-fold singularities \footnote{The isolated singularity can always be described by an affine variety \cite{ishii1997introduction}.} to engineer $\mathcal{N}=2$ SCFTs:
\begin{itemize}
\item One can use an isolated complete intersection singularity (ICIS) defined by the map $f: (\mathbb{C}^{n+3},0)\rightarrow (\mathbb{C}^n,0)$. Assume the singularity is defined by the equations $f_1=f_2=\ldots=f_n=0$, we require that these 
polynomials are quasi-homogeneous so that the weights of the coordinates $z_i$ are $w_i, i=1,\ldots, n+3$, and the degrees of $f_i$ are $d_i, i=1, \ldots, n$. The condition for the existence of a SCFT is 
\begin{equation}
\sum w_i-\sum d_i>0;
\end{equation}
\item If the ICIS has certain discrete symmetry group $G$ which preserves the canonical three form, we can form a quotient singularity using group $G$ . This will produce a large number of new SCFTs. 

\item We can also consider non-isolated singularity, and it appears that theory of class ${\cal S}$ falls in this category.

\item In general, we expect that a \textbf{rational graded Gorenstein}  isolated three-fold singularity would give us a four dimensional $\mathcal{N}=2$ SCFT.  Here graded means that the singularity should
have a $\mathbb{C}^*$ action, and Gorenstein means that there is a canonical well defined $(3,0)$ form on the singularity \cite{ishii1997introduction}, finally rational means that the weights of the $(3,0)$ form under the $\mathbb{C}^*$ action is positive. 
\end{itemize}
For these constructions, we will discuss some basic ingredients and give some illustrative examples. We leave a complete classification to the future study.

We should emphasize that the major perspective of this paper is to explore the relation between geometric singularity theory and $\mathcal{N}=2$ theory. More generally, 
one could use any 2d $(2,2)$ SCFT with $\hat{c}<2$ \footnote{We might need to put some constraints on $(c,c)$ ring.} to construct a 4d $\mathcal{N}=2$ theories \cite{Shapere:1999xr,DelZotto:2015rca} \footnote{This point has been emphasized to us by C.Vafa.}, and it would be definitely interesting to further explore along this approach.

This paper is organized as follows: section two reviews the constraints on hypersurface singularity so we can find 
a SCFT, and we further discuss various physical properties which can be extracted from geometry; 
Section three gives a complete classification for the hypersurface singularity which would define a SCFT; 
Section four discusses how to use other type of singularities to define new SCFTs; Finally a short conclusion is given in section five.

\section{Isolated hypersurface singularity and  ${\cal N}=2$ SCFT}
The dynamics of four dimensional $\mathcal{N}=2$ theory is very rich, see \cite{Tachikawa:2013kta} for a detailed review and here
we summarize some useful facts for later use. We are interested in $\mathcal{N}=2$ SCFT  so the theory has 
an $SU(2)_R\times U(1)_R$ $R$ symmetry, and the theory might also have some flavor symmetries $G$. 

$\mathcal{N}=2$ theory has an interesting moduli space of vacua which could be separated into Coulomb branch, Higgs branch, 
and mixed branches which is a direct product of a Coulomb component and a Higgs component. The IR theory on the Higgs branch 
is just a bunch of free hypermultiplets and the important question is to study its Hyperkahler metric. 
The Coulomb branch is particularly interesting: the IR theory at a generic point of the moduli space is an abelian gauge theory and 
the important task is to find out the photon couplings; there are also various singular points with 
extra massless particles where the IR theory is much more non-trivial.  Seiberg-Witten discovered that for some theories the low energy effective 
theory on the Coulomb branch could be described by a Seiberg-Witten curve fibered over the moduli space:
\begin{equation}
F(x,z,\lambda_i)=0;
\end{equation}
Here $\lambda$s are the parameters  including coupling constants, mass parameters, and expectation values for Coulomb branch 
operators. The period integral of an appropriate one form over the Riemann surface $F(x,z,\lambda)=0$ with fixed $\lambda$ determines the
low energy photon coupling. The SW curve contains a lot more information such as the central charge for BPS particles, the 
physics at the singularities on the parameter space, etc. 

So one of the most important task of studying a $\mathcal{N}=2$ SCFT is to find its SW curve for the Coulomb branch. It was soon realized 
that string theory provides the most efficient way of solving a theory. One approach is to first engineer UV theory using the type IIA brane 
configuration and then  lifting IIA configuration to a M5 brane configuration to find the SW curve \cite{Witten:1997sc}, and this might be regarded 
as an open string method.  The other approach is to first put type IIA string theory on a singularity to engineer a UV theory, and its SW solution 
is found by  mirror type IIB geometry \cite{Klemm:1996bj,Katz:1997eq}.
One of the interesting fact about the second approach is that the SW solution of some theories can only be put in a three-fold fibration $F(z_0,z_1, z_2, z_3, \lambda)=0$,
and this suggests that the most general SW solution should be a three-fold fibration rather than a curve fibration!

Instead of starting with a UV gauge theory using type IIA theory and then try to find its SW solution using type IIB mirror, we 
directly try to classify all possible three fold fibration in type IIB side which can give the SW solution of a SCFT. The task is significantly simplified 
as it appears that the most singular points namely the SCFT point on the Coulomb branch completely determine
the full SW solution, so the task of classifying a $\mathcal{N}=2$ SCFT is reduced to classify all possible three-fold singularity! In this section, we focus
on isolated hypersurafce singularity.

\subsection{Constraint on  hypersurface singularity }
Let's start with an isolated hypersurface singularity (IHS) $f:(\mathbb{C}^4,0)\rightarrow (\mathbb{C},0)$, and here we summarize the condition on 
$f$ that would give rise to a SCFT \cite{Shapere:1999xr}:
\begin{itemize}
\item $f$ has an isolated singularity at $z_i=0,~i=0,1,2,3$, which means that $f={\partial f\over \partial z_0}={\partial f\over \partial z_1}={\partial f\over \partial z_2}={\partial f\over \partial z_3}=0$ 
has a unique solution at $z_i=0$.
\item 4d $\mathcal{N}=2$ SCFT has a $U(1)_R$ symmetry, which means that the polynomial $f$ has to have a $\mathbb{C}^*$ action such that all the coordinates 
have positive weights:
\begin{equation}
f(\lambda^{q_i}z_i)=\lambda f(z_i),~~q_i>0;
\label{charge1}
\end{equation}
such polynomial is called quasi-homogenous polynomial. The $U(1)_R$ charge of 4d SCFT is proportional to this $\mathbb{C}^*$ action and 
the proportional constant will be determined later. 
\item  To get a sensible SCFT, the weights has to satisfy the following condition:
\begin{equation}
\sum q_i>1;
\end{equation}
\end{itemize}
The third condition could be  understood using string theory. Consider type IIB string theory on following background $R^{1,3}\times X^3$, where $X^3$ is an
isolated three dimensional hypersurface singularity defined by $f$. It is argued in \cite{Ooguri:1995wj,Giveon:1999zm} that in taking string coupling $g_s$ and string scale $l_s$ to zero, we get a non-trivial 
four dimensional SCFT. If we keep $l_s$ to be finite, we get a little string theory which has a holographic description on background $R^{1,3}\times R_{\phi}\times S^1\times LG(W=f)$ with proper 
orbifolding, 
here $R_\phi$ is the linear dilaton sector. To have a stable string theory, we require that the central charge $\hat{c}<2$ for the Landau-Ginzburg piece. The central charge $\hat{c}$
for the LG model defined by the superpotential $W=f$ is $4-2\sum q_i$, so $\hat{c}<2$ condition is the same as $\sum q_i>1$ condition. 

\subsection{Mini-versal deformation and Seiberg-Witten solution}
Let's start with a IHS with a $\mathbb{C}^*$ action, and assume the weights satisfy the condition $\sum q_i>1$ so we expect to get a 4d $\mathcal{N}=2$ SCFT.
The SW geometry is related to the deformation of the singularity $f$, and there is a distinguished class of deformations called mini-versal deformations which 
can be identified with the SW geometry. In the hypersurface case, the mini-versal deformation can be described easily and this is one of the most powerful 
advantage of using singularity to define a SCFT. 

Let's now describe explicitly the form of mini-versal deformations: given a quasi-homogeneous polynomial $f$ with an isolated singularity at the origin,  we can 
define the following Jacobi algebra:
\begin{equation}
J(f)={\mathbb{C}[z_0,z_1,z_2,z_3]/({\partial f\over \partial z_0},{\partial f\over \partial z_1},{\partial f\over \partial z_2},{\partial f\over \partial z_3})};
\end{equation}
here $\mathbb{C}[z_0,z_1, z_2,z_4]$ is the polynomial ring of $\mathbb{C}^4$, and
the above algebra has finite dimension $\mu$ since $f$ has an isolated singularity. 
Let's take $\phi_1(z),\ldots, \phi_\mu(z)$ as the monomial basis of the above Jacobi algebra, then the Seiberg-Witten solution is given by:
\begin{equation}
F(z,\lambda)=f(z_0,z_1,z_2,z_3)+\sum_{i=1}^\mu \lambda_\alpha\phi_\alpha(z)=0;
\end{equation}
There is a also an canonically defined SW differential:
\begin{equation}
\Omega= {dz_0\wedge dz_1 \wedge dz_2\wedge dz_3 \over dF}.
\end{equation}

Let's use $z^\alpha$ to denote one basis vector of the Jacobi algebra, and $u_\alpha$ to denote the coefficient before the monomial in SW geometry. 
The coefficient $u_{\alpha}$ corresponds to physical parameter
 that would deform Coulomb branch of a $\mathcal{N}=2$ SCFT. 

The scaling dimension of $\Delta(u_\alpha)$ can be easily found from the charge $Q^\alpha$ of $z^\alpha$  
under the $\mathbb{C}^*$ action presented in $\ref{charge1}$. The $U(1)_R$ charge is proportional to the charge under 
$\mathbb{C}^*$ action, and so the scaling dimension of an operator on Coulomb branch is also proportional to its $\mathbb{C}^*$ charge
 as those operators are chiral primary. Based on above fact, we assume that 
 the scaling dimension $\Delta$ of an operator is proportional to the $\mathbb{C}^*$ charge: $\Delta=\delta Q_\alpha$.  $\delta$ can be found 
using the following condition: the canonical differential $\Omega$
 has $Q$ charge $(\sum q_i-1)$, and $\Omega$ is required to have scaling dimension $1$ as the integration of this three form on three cycles would give the mass for the BPS particle, so we have 
$(\sum_{i=0}^{3} q_i-1) \delta=1$, and we find 
\begin{equation}
\delta= {1\over \sum_{i=0}^{3} q_i-1 }={2\over 2-\hat{c}}.
\end{equation}
with $\hat{c}=4-2\sum_{i=0}^{3} q_i$, here $\hat{c}$ is the normalized central charge for the two dimensional $(2,2)$ Landau-Ginzburg model defined by $f$. 
For a deformation $u_{\alpha} z^{\alpha}$, the $\mathbb{C}^*$ charge of $u_\alpha$ is $1-Q_{\alpha}$ so that the total weights of this deformation term is one, then the scaling dimension of $u_\alpha$ is 
\begin{equation}
[u_\alpha]=(1-Q_\alpha)\delta={1-Q_\alpha \over \sum q_i-1}={2(1-Q_{\alpha})\over 2-\hat{c}}.
\end{equation}
Notice that generically there could be deformations whose coefficients have  negative scaling dimensions.

The Jacobi algebra plays a crucial role in defining the mini-versal deformations, and the Poincare polynomial of the algebra $J(f)$ 
can be computed using the weights information. To define this polynomial, we introduce the following equivalent $\mathbb{C}^*$ action
\begin{equation}
f(\lambda^{w_0}z_0,\ldots, \lambda^{w_3} z_3)=\lambda^df(x_0,\ldots, x_n).
\end{equation}
Now the weights $w_i$ and degree $d$ are all positive integers, and one can recover 
previous weights using the relation $q_i={w_i\over d}$. Define the Poincare polynomial based on $\mathbb{C}^*$ action
\begin{equation}
P(t)=\sum_{\alpha} (\text{dim} H_{\alpha}) t^\alpha;
\end{equation}
Here $\alpha$ is the weight of $\mathbb{C}^*$ action and $\text{dim} H_{\alpha}$ is the dimension of the subspace with charge $\alpha$. The Poincare polynomial 
is \cite{arnold2012singularities}:
\begin{equation}
P(t)=\prod_{i=0}^3{1-t^{d-w_i}\over 1-t^{w_i}}.
\end{equation}
The dimension of Jacobi algebra is then 
\begin{equation}
\mu=\prod_{i=0}^3 ({d\over w_i} -1)=\prod_{i=0}^3 ({1\over q_i} -1);
\end{equation}
and the maximal degree of the monomial basis vector is 
\begin{equation}
d_{max}=4d-2\sum w_i.
\end{equation}  
 The minimal degree is obviously zero.  

There are several simple comments on the possible spectrum:
\begin{itemize}  
\item The deformations are paired except for the deformation with scaling dimension $1$, and these pairs satisfy the following condition
\begin{equation}
[m]+[u]=2;
\end{equation} 
This is required by $\mathcal{N}=2$ supersymmetry. 
\item There is a unique operator with highest scaling dimension which is given by the constant deformation $\lambda\cdot1$, and its scaling dimension is
\begin{equation}
[u]_{max}=2/(2-\hat{c}).
\end{equation}
\end{itemize}

The spectrum of operators with positive scaling dimensions could be classified according to their scaling dimensions:
\begin{itemize}
\item $[u]> 1$: Coulomb branch operators, and this happens if $Q_\alpha<{\hat{c}\over 2}$. Among them, we call an operator with scaling dimension $1<[u]<2$ relevant operator.
\item The deformation $[u]=1$ is called mass parameter, this happens if $Q_\alpha={\hat{c}\over2}$. 
\item The deformation $ 0\leq[u]<1$ are called coupling constants, and this happens if $Q_\alpha >{\hat{c}\over2}$; An operator with scaling dimension $[u]=0$ is called 
exact marginal deformation, and this happens if $Q_\alpha=1$. 
\end{itemize}
In the following we use $r$ to denote the number of deformations with scaling dimension bigger than one, and $r$ is called dimension of the Coulomb branch; We 
also use $f$ to denote the number of deformations with scaling dimension one which are mass parameters. The rank of the charge lattice of the theory is then 
\begin{equation}
R=2r+f=\mu;
\end{equation}
and the relation between the dimension of Jacobi algebra and the rank of the charge lattice is due to the paring of  mini-versal deformations. 

Let's make some comments on the spectrum of the theory engineered using hypersurface singularity: 
first, all kinds of possibilities can happen: it is possible to have mass parameters or not to have mass parameters;
it is possible to have the exact marginal deformations or not have exact marginal deformations; the theory can have relevant or not have relevant operators. 
If there is an exact marginal deformation in our theory, it is possible 
to find a weakly coupled gauge theory description and it is interesting to study the S duality behavior of the theory;  If the theory has mass parameters, it is interesting to study whether there
is non-abelian flavor symmetry, etc. 

One could try to classify $\mathcal{N}=2$ SCFT based on the property of the spectrum on the Coulomb branch:
\begin{itemize} 
\item One could classify the theory based on the dimension of the Coulomb branch (the number of operators on Coulomb branch with scaling 
dimension bigger than one). An attempt of trying to classify rank one theory based on Kodaira's classification of singular elliptic fibre has been 
given in \cite{Argyres:2015ffa}.  Some of them can be realized using IHS:
\begin{align}
&H_0:  f=z_0^2+z_1^2+z_2^2+z_3^{3}, \nonumber\\
&H_1: f=z_0^2+z_1^2+z_2^2+z_3^{4}, \nonumber \\
&H_2:f=z_0^2+z_1^2+z_2^3+z_3^{3}.
\end{align}
It is interesting to classify all rank one, rank two theories which can be realized by IHS. 

\item One could also
classify the theory based on the number of irrelevant and exact marginal deformations of the mini-versal deformations. For SCFT without any marginal and irrelevant deformations in mini-versal deformation, one has the following 
ADE sequences and their SW solutions:
\begin{align}
& A_k:~~f=z_0^2+z_1^2+z_2^2+z_3^{k+1},\nonumber\\
&~~F(z,\lambda)=f+\lambda_1+\lambda_2z_3+\ldots+\lambda_{k}z_3^{k-1}; \nonumber\\
& D_k:~~f=z_0^2+z_1^2+z_2^{k-1}+z_2z_3^2, \nonumber\\
&~~F(z,\lambda)=f+\lambda_1+\lambda_2z_3+\lambda_3z_2+\ldots+\lambda_kz_2^{k-2}; \nonumber\\
&E_6:~~f=z_0^2+z_1^2+z_2^{3}+z_3^4, \nonumber \\
&~~F(z,\lambda)=f+\lambda_1+\lambda_2 z_2+\lambda_3 z_3+\lambda_4 z_3^2+\lambda_5 z_2 z_3+\lambda_6 z_2 z_3^2; \nonumber\\
&E_7:~~f=z_0^2+z_1^2+z_2^{3}+z_2z_3^3, \nonumber \\
&~~F(z,\lambda)=f+\lambda_1+\lambda_2 z_2+\lambda_3 z_3+\lambda_4 z_3^2+\lambda_5z_3^3+\lambda_6 z_3z_2^2+\lambda_7 z_2 z_3; \nonumber\\
&E_8:~~f=z_0^2+z_1^2+z_2^{3}+z_3^5. \nonumber \\ 
&~~F(z,\lambda)=f+\lambda_1+\lambda_2 z_2+\lambda_3 z_3+\lambda_4 z_3^2+\lambda_5z_3^3+\lambda_6 z_2 z_3+\lambda_7 z_2 z_3^2+\lambda_8 z_2z_3^3;
\end{align}
These  ADE Argyres-Douglas SCFTs are found as the maximal singular point at the corresponding pure $\mathcal{N}=2$ ADE gauge theory \cite{Eguchi:1996vu}. 
The next class would be the SCFT whose spectrum consists of operators with \textbf{non-negative} scaling dimension. Those theories 
are actually called complete theory in \cite{Cecotti:2011rv}.  In general, we would try to classify the theory by the number of deformations with non-positive scaling dimensions,
and we denote this number as $m$. This number is actually the modality in singularity theory \cite{arnold2012singularities}. 
\end{itemize}
 
\textbf{Example}: Let's consider the singularity $f=z_0^a+z_1^b+z_2^c+z_3^d$ with the constraint ${1\over a}+{1\over b}+{1\over c}+{1\over d}>1$, and the Milnor number is  $\mu=(a-1)(b-1)(c-1)(d-1)$. 
The relation from Jacobi ideal generated by ${\partial f\over \partial z_i}=0$ is simply 
\begin{equation}
z_0^{a-1}=0,~~z_1^{b-1}=0,~~z_1^{c-1}=0,~~z_1^{d-1}=0.
\end{equation}
So the corresponding Jacobi algebra has the following monomial basis:
\begin{equation}
z_0^{\alpha}z_1^{\beta}z_2^{\gamma}z_3^{\delta},~~~~0\leq \alpha\leq a-2,~~0\leq \beta\leq b-2,~~0\leq \gamma\leq c-2,~~0\leq \delta\leq d-2,~~
\end{equation}
The total dimension of this algebra is $(a-1)(b-1)(c-1)(d-1)$. 
The scaling dimension of the coefficient for the above deformation is 
\begin{equation}
[u]_{\alpha\beta\gamma\delta}={1-({\alpha\over a}+{\beta\over b}+{\gamma\over c}+{\delta\over d})\over {1\over a}+{1\over b}+{1\over c}+{1\over d}-1}.
\end{equation}

\subsection{Discriminant locus, bifurcation diagram and extra massless particles}
\begin{figure}[h]
    \centering
    \includegraphics[width=3.0in]{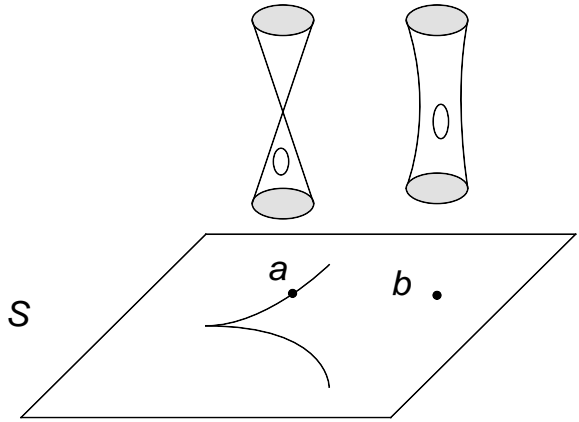}
    \caption{S: the Coulomb branch moduli space for $\mathcal{N}=2$ theory. Here the curve on S means the discriminate locus; a is a point on which the Milnor fiber becomes singular, b
    is a point where the Milnor fiber is smooth.}
    \label{discrimi}
\end{figure}

Let $f:(\mathbb{C}^4,0)\rightarrow (\mathbb{C},0)$ be a IHS defining a $\mathcal{N}=2$ SCFT , and let's denote the basis of the Jacobi algebra as $\phi_\alpha(z),\alpha=1,\ldots,\mu$; The miniversal deformation of 
$f$ is written as
\begin{equation}
F(z,\lambda)=f(z)+\sum_{i=0}^{\mu-1}\lambda_\alpha \phi_\alpha(z)=0;
\end{equation}  
here we take $\phi_0(z)=1$. The parameters $\lambda_i$ control the deformation of SCFT in Coulomb branch (some of $\lambda_i$ are coupling constants, and we treat them 
at the same footing as the operator with dimension larger than one as they all could change the low energy effective theory). The subspace $\Sigma$ on $S$ at which $F(z,\lambda)$ is singular 
is called as the discriminant locus. It is known that $\Sigma$ is an irreducible hypersurface and its multiplicity is equal to the Milnor number $\mu$, 
see figure. \ref{discrimi}. On the space $S^{'}=S/ \Sigma$, the fibre is non-singular, and its only non-vanishing homology class is $H^0$ and $H^3$. An amazing fact 
about the non-singular fibre is that it is a bouquet of $\mu$ $S^3$, so one can get BPS particle by wrapping $D3$ branes on these $S^3$s. 

At the discriminant locus $\Sigma$, some of the $S^3$ have  vanishing area, and 
there are extra massless particles coming from D3 brane wrapping on vanishing cycle. The IR theory would be quite different at those special points. 
The study of the the structure of discriminant locus is an important part in understanding the interesting dynamics of a $\mathcal{N}=2$ theory. In singularity theory, the structure 
of the discriminant locus is called bifurcation diagram which is also a quite important subject. The result from singularity literature could definitely teach us about the
dynamics of $\mathcal{N}=2$ theory.

\begin{figure}[h]
    \centering
    \includegraphics[width=2.0in]{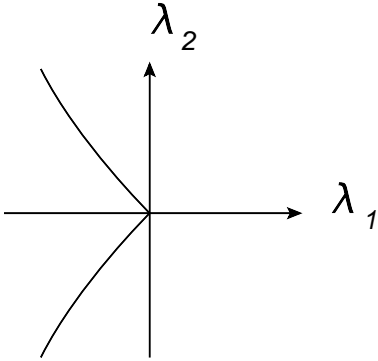}
    \caption{The discriminate locus of $A_2$ SCFT, here the parameters $\lambda_1, \lambda_2$ are taken to be real.}
    \label{bifurcation}
\end{figure}

\textbf{Example}: Let's consider  $A_2$ AD theory defined by the IHS $f=z_0^2+z_1^2+z_2^2+z_3^3$, and the mini-versal deformation of this 
singularity is $z_0^2+z_1^2+z_2^2+z_3^3+\lambda_1 z_3+\lambda_2=0$. The quadratic terms will not change the discriminant locus, and we can focus on the geometry 
$z_3^3+\lambda_1 z_3+\lambda_2=0$, and the discriminant locus is given by the 
equation 
\begin{equation}
\lambda_1^3+{27\over 4}\lambda_2^2=0. 
\end{equation}
which is simply the condition that the roots of the above cubic equations are degenerate, see illustration in figure. \ref{bifurcation}.

\subsection{Period integral and low energy effective action}

Let's denote the moduli space as $S$, and the discriminant locus as $\Sigma$. The hypersurface $F(z,\lambda)=0$ 
over fixed $\lambda$ is smooth away from $\Sigma$, see figure. \ref{discrimi}. 
This is called central Milnor fibration whose middle homology $H^3(F_\lambda, C)$ has dimension $\mu$, which is called the Milnor number.

We would like to understand the low energy spectrum from type IIB string theory point of view.  Type IIB string theory has a self-dual 
four form $A_{\mu\nu\rho\sigma}$, and if we comactify IIB string theory on a compact Calabi-Yau manifold, we get  $h^{2,1}$ vector multiplet. 
This is due to the fact that there is a hodge structure on the manifold such that $h^{2,1}=h^{1,2}$, and the self-duality 
condition could be solved automatically by this Hodge structure. In our current context, there is no standard Hodge structure, but 
one can define a mixed Hodge structure \cite{arnol?d2012singularities}. We do not explain the detail about mixed hodge structure here, and we just point out a crucial fact that the paired hodge number
is nothing but $r$ which is equal to the dimension of the coulomb branch (number of operators with scaling dimension bigger than one). The unpaired 
Hodge number is equal to the number of mass parameters. With this Hodge structure, we now have  $r$ vector multiplets in four dimension from compactification of 
self-dual four form. 
From $\mathcal{N}=2$ field theory point of view, at a non-singular point of the moduli space $S$, the IR theory is described by a  $U(1)^r$  gauge theory. 

The low energy coupling for the photons is related to the period integral associated with the $\mathcal{N}=2$ geometry. Let's discuss this period integral 
in more detail: take the vector space of middle homology of 
the Milnor fibration, one can get a homology bundle; Similarly, take the vector space of middle cohomology, one can a cohomology bundle. The period 
integral is basically a pairing between the cohomology bundle and homology bundle. 
Let's take a continuous integral basis $\delta_1(\lambda),\ldots, \delta_\mu(\lambda)$ of the middle homology group of Milnor fibration, and one can  form the following period integral:
\begin{equation}
\lambda\rightarrow (\int_{\delta_1(\lambda)} \Omega,\ldots,\int_{\delta_\mu(\lambda)} \Omega).
\end{equation}
Here $\Omega={dz_0\wedge dz_1\wedge dz_2\wedge dz_3\over d F(z,\lambda)}$ is the canonical holomorphic three form on the fibre.  
The low energy effective action of the theory could be read from the information of these period integrals.  There is an intersection form (possibly degenerate)
on the middle  homology, and we can choose a basis such that they have the following intersection form:
\begin{align}
&A^i\cdot B_j=\delta_{ij},~~i=1,2,\ldots r; \nonumber\\
& L_\alpha\cdot A^i=0,~~L_\alpha\cdot B_i=0,~~ L_\alpha\cdot L_{\beta}=0,~\alpha=1,\ldots, f.\nonumber\\
\end{align}
Locally on the moduli space, We might identify the periods as 
the electric, magnetic and the mass coordinates on the moduli space: 
\begin{align}
&a^i(\lambda)=\int_{A^i(\lambda)} \Omega,~~~a^D_i(\lambda)=\int_{B_i(\lambda)} \Omega, \nonumber\\
&m_\alpha(\lambda)=\int_{L_{\alpha}(\lambda)}\Omega.
\end{align}
In this basis, the photon coupling would be 
\begin{equation}
\tau^{ij}={{d a_{i}^D \over du_k} {d  u_k \over d a^i  }},~~i,j=1,\ldots,r.
\end{equation}
Here $u_i$ are the operators with scaling dimension bigger than one. 
For IHS and its deformation, these period integral has important property such as holomorphy dependence on the parameters, etc.

\subsection{Monodromy, vanishing cycles and BPS quiver }
The physical interpretation of the singularity on the base of the miniversal deformation of IHS $f$ is that there are 
extra massless particles \cite{Seiberg:1994rs}. There massless particles are from the stable massive BPS particles on 
the non-singular locus of the moduli space. From type IIB string point of view, 
the BPS particle comes from D3 brane wrapping on special Lagrangian three cycle of the geometry $F(z,\lambda)=0$, and therefore
the existence of extra particle is related to the vanishing cycle on the singularity. 
 
\begin{figure}
    \centering
    \includegraphics[width=3.0in]{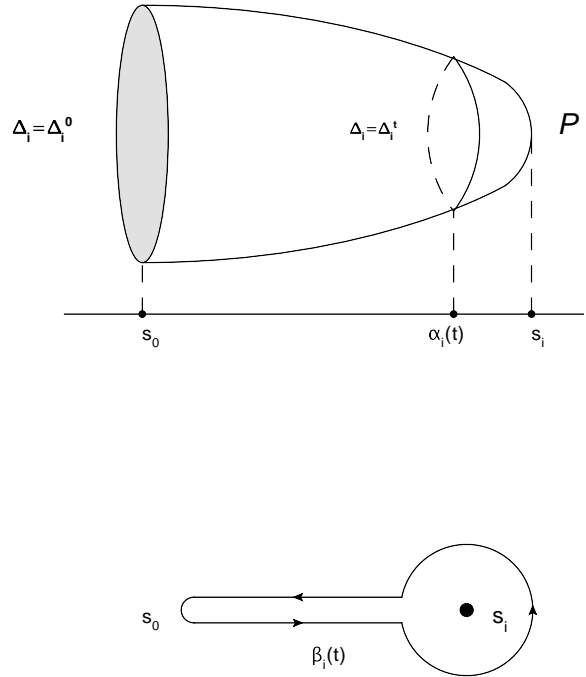}
    \caption{Up: The three cycle becomes vanishing cycle in approaching the singular point $s_i$. Bottom: A path around  the singular point on the moduli space.}
    \label{vanish}
\end{figure}

To make this picture more precise, let's focus on co-dimensional one singularity on $S$. 
Let's fix a point $s_0$ and form a path $\alpha(t)$ connecting $s_0$ and a co-dimension one singularity. 
Certain three cycle becomes vanishing  along the path , see figure. \ref{vanish}. One get an extra 
massless particles from D3 brane wrapping on this vanishing cycle. 

There is a monodromy action 
on the homology of the fibration which is associated with the loop $\beta_i$ around one singularity $s_i$, see figure. \ref{vanish}. The monodromy action is given by the Picard-Lefschetz  transformation:
\begin{equation}
T_i(x)=x+<x,\Delta_i>\Delta_i,
\end{equation}
here $x$ is an element of integral homology and 
$\Delta_i$ is the vanishing cycle, and $<x,\Delta_i>$ is the intersection number between $x$ and $\Delta_i$.  Considering the period integral, we have the following monodromy transformation
\begin{equation}
\int_x \Omega\rightarrow \int_{x+<x,\Delta_i>\Delta_i}\Omega=\int_x\Omega+<x,\Delta_i>\int_{\Delta_i}\Omega;
\end{equation}
These are just the electric-magnetic duality actions on the low energy effective $U(1)$ theories.

\begin{figure}
    \centering
    \includegraphics[width=2.0in]{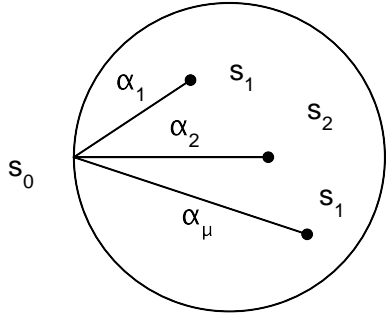}
    \caption{A set of loops around various $A_1$ singularities, and these generate the $\pi_1$ of the base of the fibration.}
    \label{loop}
\end{figure}

More generally, for IHS, one can turn on deformations such that there are a total of $\mu$ codimensional one singularities. These
singularities are the so-called  $A_1$ singularity around which the 
equation can be denoted as 
\begin{equation}
z_0^{'2}+z_1^{'2}+z_2^{'2}+z_3^{'2}=\lambda_0;
\end{equation}
here $z_i^{'}$ is the local coordinates around the singularity. This kind of deformation is called Morsification of the singularity. 
Let's fix a point $s_0$ which is sufficiently closed to origin. One can choose a line L close to the origin which meets discriminant $\Sigma$ in $\mu$ points $s_1, \ldots, s_\mu$.  
We can take a non self-intersecting path $\alpha_i$ from a fixed point $s_0$ to $s_i$, see figure. \ref{loop}. The ordering of $s_i$ and $\alpha_i$ is given by rule  such that 
they start from $s_0$ and counting clockwise.  Since each path gives us a vanishing cycle, we therefore get a ordered basis $\Delta_i, i=1,\ldots,\mu$ of the middle homology of the Milnor fibration. 
Given these paths, one can form a bunch of loops which generate $\pi_1(S/\Sigma)$ and  there is a total monodromy 
\begin{equation}
T=T_1T_2\ldots T_\mu. 
\end{equation}
This monodromy matrix plays an important role in characterizing the singularity. 

The above product of the monodromy is related to the total monodromy of singularity, which is defined using a so-called Milnor fibration.
The Milnor fibration is defined as follows: Let's take a disc $D_\delta=\{t\in C; 0<|t|<\delta$, and the Milnor fibration associated with 
a IHS is 
\begin{equation}
\Psi: \{|z_0|^2+|z_1|^2+|z_2|^2+|z_3|^2=\epsilon^2\} \bigcap f^{-1}(D_\delta)\rightarrow D_\delta.
\end{equation}
This map is a topologically locally trivial fibration for any $\epsilon>>\delta>0$. It is not hard to see that this fibration is the fibration 
around the origin of the Coulomb branch by turning on constant deformation. Choosing a small cycle around the origin, then 
the monodromy 
acts on the homology of the Milnor fibration as 
\begin{equation}
T:~H_3(\Psi^{-1}(t),C)\rightarrow H_3(\Psi^{-1}(t),C).
\end{equation} 
This monodromy group is just the one defined earlier using the Picard-Lefshertz transformations around co-dimensional one singularity. 
There are several important features about the monodromy:
\begin{itemize}
\item The eigenvalues of the monodromy matrix is a root of unity, and the eigenvalues are 
related to basis of Jacobi algebra. Consider a monomial basis $z^\alpha$, then the eigenvalue associated with it is $\exp(2\pi i l_i)$ with 
\begin{equation}
l_i=\sum_{j=0}^3 (n_j+1)q_j-1,
\end{equation}
here $n_j$ is the exponent of $z^{\alpha}$. It is easy to check that the number of  unity eigenvalues is equal to the number of mass parameters by noting the following two facts: a: $l_i=1$ 
gives us a mass parameter; b: $0<l_i<2$. 
\item The monodromy matrix satisfies the condition $(T^N-I)^k=0$ for some $N$ and $k$, namely $(T^N-I)$ is an nilpotent matrix. 
\end{itemize} 

\begin{figure}
    \centering
    \includegraphics[width=5.0in]{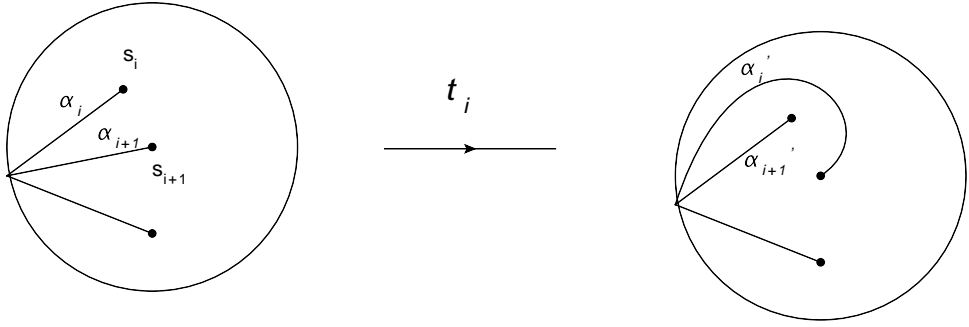}
    \caption{Braiding operation which changes the basis of vanishing cycle.}
    \label{change}
\end{figure}

Those vanishing cycles are the distinguished basis of the middle homology of the Milnor fibration. 
The important data is the intersection form on the set $(\Delta_1, \ldots, \Delta_\mu)$, and one can form an antisymmetric intersection matrix as follows:
\begin{align}
& I_{\Delta}=<\Delta_i,\Delta_j>_{1\leq i,j\leq \mu}, i\neq j \nonumber\\
&<\Delta_i,\Delta_i>=0
\end{align}
These intersection form depends on the choices of paths and is not unique. There are two operations which can be used to change the basis. 
The first operation is to change the orientation of the cycle:
\begin{equation}
s_i:(\Delta_1, \ldots, \Delta_i,\ldots, \Delta_\mu)\rightarrow (\Delta_1, \ldots, -\Delta_i,\ldots, \Delta_\mu)
\end{equation}
The second transformation is more nontrivial, and it acts the path by the so-called braiding, and see figure. \ref{change} for how the path is changed. The basis is 
changed as 
\begin{equation}
t_i:(\Delta_1, \ldots, \Delta_i,\Delta_{i+1}\ldots, \Delta_\mu)\rightarrow (\Delta_1, \ldots, \Delta_{i+1}+<\Delta_{i+1},\Delta_{i}>\Delta_i,\Delta_i,\ldots, \Delta_\mu)
\end{equation}
It is straightforward to derive the change of the intersection form under this change of basis. 

We now conjecture that the intersection form of vanishing cycle might be identified with the BPS quiver.  Let's choose the distinguished basis and perform 
the period integral 
\begin{equation}
\lambda\rightarrow (\int_{\Delta_1(\lambda)} \Omega,\ldots,\int_{\Delta_\mu(\lambda)} \Omega)=(a_1(\lambda),\ldots, a_\mu(t)).
\end{equation}
Consider a special Lagrangian three cycle in Homology class $L=n_1 \Delta_1+\ldots n_\mu \Delta_\mu$, then 
the central charge of it is given by the integral of $(3,0)$ form on $L$
\begin{equation}
Z=\int_L \Omega=\int_{\sum n_i \Delta_i}\Omega=\sum n_i a_i.
\end{equation}
Due to special Lagrangian condition, the mass of this particle is equal to the absolute value of the central charge. 
The task of counting BPS spectrum is to determine which $n_i$ are allowed in certain region of the moduli space, and 
study the wall crossing behavior of various BPS particles. The BPS quiver plays a crucial role in finding the spectrum, 
so knowing the BPS quiver is a first step. For many known theories in this class, the BPS quiver is actually the intersection form of the vanishing cycle \cite{Cecotti:2010fi,Cecotti:2011gu, DelZotto:2011an,}, and we would like to 
conjecture that this fact is true for all the theories defined by IHS, other evidences will be given elsewhere.  

We hope that the combination of geometric interpretation of the BPS quiver and the representation 
theory of the quiver would give us more information about the full BPS spectrum of these theories.

\textbf{Example}: Consider $A_2$ SCFT, the Milnor number is 2, and the intersection form associated with vanishing cycle is $<\Delta_1, \Delta_2>=1$.
The monodromy group can be easily found using Picard-Lefschetz transformation: 
\begin{align}
&T_1=\left(\begin{array}{cc}
1&0\\
1&1\\
\end{array}\right),
\nonumber \\
& T_2=\left(\begin{array}{cc}
1&-1\\
0&1\\
\end{array}\right),~~~\nonumber \\    
& T=T_1T_2=\left(\begin{array}{cc}
1&-1\\
1&0\\
\end{array}\right).
\end{align}

\subsection{Central charge $a$ and $c$}
For four dimensional conformal field theory, there are two important central charges $a,c$ which measure the degree 
of freedom of a theory, usually it is pretty difficult to compute those quantities. For $\mathcal{N}=2$ SCFTs , the central charges 
can be computed using various methods such as free theory limit, three dimensional mirror, etc. 
Here we are going to give
an extremely simple formula for the central charges for all the theory defined using IHS. 

The formula which we are going to use is the one discovered in  \cite{Shapere:2008zf}:
\begin{equation}
a={R(A)\over 4}+{R(B)\over6}+{5 r\over 24}+{h\over 24},~~c={R(B)\over 3}+{r\over 6}+{h\over 12},
\label{central}
\end{equation} 
Here $R(A)$ is given by the Coulomb branch spectrum consists of operators with scaling dimension bigger than one:
 \begin{equation}
 R(A)=\sum_i([u_i]-1),
 \end{equation}
 and $r,h$ are the number of free vector multiplets and free hypermultiplets at the generic point of Coulomb branch. 
Notice that there is no free hypermultiplet for the class of theories we considered in this paper, which can be verified by the fact that there is 
only non-vanishing middle homology class. The above remarkable formula is derived using the topological twisting of a $\mathcal{N}=2$ theory \cite{Witten:1995gf,Shapere:2008zf}. 
 
 For theories defined by IHS, $R(A)$ and $r$ can be 
 easily found from the mini-versal deformation; Using the deformation pattern of singularity theory, we are going to give a 
simple formula for $R(B)$.  $R(B)$ is related to the co-dimensional one singularity on which there is an extra massless 
 hypermultiplet, and is given by the R charge of local coordinate near such singularity ${1\over 4}R(\delta z)$. $R(B)$ is the sum of  contribution from all co-dimensional one singularities:
 \begin{equation}
 R(B)=\mu {1\over 4}R(\delta z);
 \end{equation}
 Here $\mu$ is the number of co-dimensional one singularity, $R(\delta z)$ is the $R$ charge of the local coordinates.   
The crucial fact for us is that there are a total of $\mu$ co-dimension one  $A_1$ singularity at the Coulomb branch \cite{looijenga1984isolated}. 
Near the singularity,  the three fold takes the following simple form:
 \begin{equation}
 \delta z_0^2+ \delta z_1^2+\delta z_2^2+\delta z_3^2=\lambda_0.
 \end{equation}
We would like to know the $U(1)_R$ charge of the local coordinates $\delta z_i$. 
It is easy to see that this singularity is the $A_1$ singularity which represents a massless 
hypermultiplet.  The only deformations are constant deformation whose scaling dimension is 
\begin{equation}
\Delta(\lambda_0)={2\over 2-\hat{c}}, 
\end{equation}
 so the local coordinates $\delta z_i$ has scaling dimension $[\delta z_i]={1\over 2}\Delta(\lambda_0)$ and therefore its $U(1)_R$ charge 
 is $R(\delta z_i)=2[\delta z_i]={2\over 2-\hat{c}}$, so $R(B)$ is equal to 
 \begin{equation}
 R(B)={\mu \over 2(2-\hat{c})}={\mu \over 4(\sum q_i-1)}={1\over 4} \mu [u]_{max}.  
 \end{equation}
 Here $[u]_{\max}$ is the operator with maximal scaling dimension on the Coulomb branch.
 
 \textbf{Example I}: Let's check our formula for the simplest theory defined by $f=z_0^2+z_1^2+z_2^2+z_3^2$, and this is the free hypermultiplet. The "Coulomb branch" 
 is described by the formula $F=z_0^2+z_1^2+z_2^2+z_3^2+\lambda$, and $\lambda$ has scaling dimension one, so $R(A)=0$. There is only one $A_1$ singularity on 
 the origin, so $R(B)={1\over 4}$. We have $r=h=0$, and using the formula in \ref{central},  we find the central charge 
 \begin{equation}
 a={1\over 24},~~c={1\over 12}.
 \end{equation} 
 This is nothing but the central charge for a free hypermultiplet.

 \textbf{Example II}: 
 Let's check our formula for a more complicated example. The singularity is given by $f=z_0^3+z_1^3+z_2^3+z_3^{3k}$, and the Milnor number is $\mu=8(3k-1)$. Using singularity and its
 mini-versal deformation, we find: 
 \begin{equation}
 R(A)=-6 k + 12 k^2,~~~~R(B)={6k(3k-1)},~~~~r=12k-7,
 \end{equation} 
and the central charge is given by 
\begin{equation}
a=6k^2-\frac{35}{24},~~~c= 6 k^2-\frac{7}{6}.
\end{equation}

\begin{figure}
    \centering
    \includegraphics[width=3.0in]{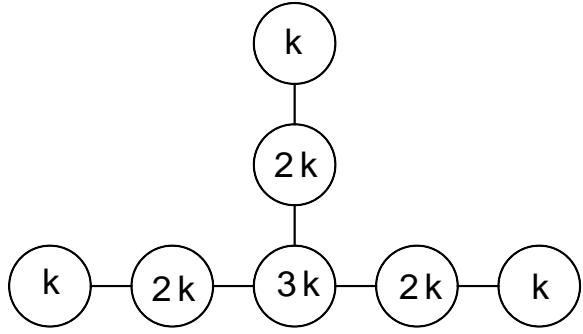}
    \caption{Lagrangian description for the singularity $f=z_0^3+z_1^3+z_2^3+z_3^{3k}$.}
    \label{charge}
\end{figure}

It is known \cite{DelZotto:2015rca} that this is affine $E_6$ quiver gauge theory, see figure. \ref{charge} for the quiver.
This theory has a Lagrangian description and its central charge can be easily found using the following formula
 \begin{equation}
 a={n_h\over 24}+{5 n_v\over 24},~~c={n_h\over 12}+{n_v\over 6}.
 \end{equation}
 Substitute $n_h=24 k^2,~~n_v=24k^2-7$ into above formula, we find
 \begin{equation}
a= 6 k^2-\frac{35}{24},~~c= 6 k^2-\frac{7}{6};
 \end{equation}
which agrees with the result derived above using totally different method. 

\subsection{Exact marginal deformations, duality group and moduli of singularity}
If we can find a monomial $z^{\alpha}$  with weights $1$ in the monomial basis of Jacobi algebra, there is an exact marginal deformation in our theory. 
The space of exact marginal deformations is therefore identified with the following family of singularity. 
\begin{equation}
f(z_0,z_1,z_2,z_3)+\sum \lambda_{\alpha} z^{\alpha}=0,
\end{equation}
Here the sum is over the monomials with weight 1. This family of deformations have a distinguished feature that the Milnor number 
is constant along this deformation (notice that the irrelevant deformation also gives rise to $\mu$ constant deformation). 
The coupling constant space is therefore identified as $\mathbb{C}^p$, where $p$ is the number of weight one deformations. 
However, these points do not give  singularity with different complex structures as two different number $\lambda$ and $\lambda^{'}$ could give 
bi-holomorphic equivalent singularity. 
An important question is to determine an invariant function so that two isomorphic coupling constant would give 
the same answer (this is the analog of the J invariant for the elliptic curve). After finding the invariant, one can find out the modular group $G$  and 
the the space of exact marginal deformations is therefore $\mathbb{C}^p/G$. This $G$ is nothing but the duality group.

On the other hand, the equation defines a weighted homogeneous variety if we quotient the above hypersurface by the defining $\mathbb{C}^*$ action, therefore 
the space of exact marginal deformations are the space of complex structure deformations of the corresponding variety. This is a generalization of 
class ${\cal S}$ theory in which the space of exact marginal deformations are identified with the complex structure moduli of curves, here
we would find the complex structure moduli space of surfaces.

Whenever there is 
an exact marginal deformation, the theory should have a weakly coupled gauge theory description and the exact marginal deformation could 
be identified with the gauge coupling. It appears that 
the mirror symmetry for $f$ plays a important role in finding the gauge theory description.  
The detailed study of space of exact marginal deformation and 
the gauge theory description  of these theories will be left to a separate publication, see also \cite{Buican:2014hfa, DelZotto:2015rca, Cecotti:2015hca} for the studies of gauge theory descriptions of some singularities.

\textbf{Example I}: Let's consider the singularity $f=z_0^2+z_1^3+z_2^3+z_3^3$, and there is one weight one deformation $z_1 z_2 z_3$. 
So we have a family of singularities $F=z_0^2+z_1^3+z_2^3+z_3^3+\lambda z_1 z_2 z_3$ with the constraint $\lambda^3+27\neq0 $ so that 
there is an isolated singularity at the origin. Let's ignore $z_0$ term as 
it does not contribute anything to our problem. The projective variety defined by $z_1^3+z_2^3+z_3^3=0$ is a torus, and $\lambda$ is the complex 
structure of this variety. The invariant for this singularity has been calculated by Saito \cite{saito1974einfach}:
\begin{equation}
J(\lambda)=-{\lambda^3(\lambda^3-216)^3\over 1728(t^3+27)};
\end{equation}

\textbf{Example II}: Let's consider following singularities:
\begin{align}
& z_0^3+z_1^3+z_2^3+z_3^3=0, \nonumber\\
& z_0^2+z_2^4+z_3^4+z_3^4=0, \nonumber\\
& z_0^2+z_2^3+z_3^6+z_3^6=0.
\end{align}
These IHS define quiver gauge theory of affine $E_6, E_7, E_8$ shape respectively \cite{Katz:1997eq,DelZotto:2015rca}. The number of exact marginal deformations are $4,6,8$ respectively. 
The corresponding weighted projective variety 
is the smooth Del Pezzo surface $dP_6, dP_7, dP_8$ respectively. So the space of gauge coupling is identified with the complex structure 
of those algebraic surfaces. 

\subsection{RG webs and adjacency of singularity}
If there is a relevant deformation in our spectrum, we can turn on this deformation and flow to other theories. Such RG flow has been studied in \cite{Xie:2013jc} for some theories,
here we will give a more general illustration. 

One can turn on a relevant deformation of a $\mathcal{N}=2$ SCFT and flow to a new fixed point in the infrared, i.e. one can turn on the deformation \cite{Argyres:1995xn}:
\begin{equation}
\int d^4 \theta~{\langle v\rangle\over \mu^{\sigma}} U,
\end{equation}
where $\langle v\rangle$ is the expectation value of certain operator with dimension $1$. The coupling constant is identified as $m={\langle v\rangle\over \mu^{\sigma}}$ with scaling dimension 
\begin{equation}
[m]=1-\sigma,
\end{equation}
and the operator $U$ has dimension $[U]=1+\sigma$. $U$ is an irrelevant operator when $[U]>2$, marginal when $[U]=2$, and relevant when $[U]<2$. Here  $U$ is operator 
for which we can turn on its expectation value which parameterizes the Coulomb branch. The scaling dimensions of $m$ and $U$ satisfy the relation $[m]+[U]=2$. We would like 
to study the IR theory in the limit $\mu\rightarrow 0$. 

Let's now consider a SCFT defined by a IHS.  If we turn on the relevant deformation $m$ (without turning on 
the expectation value of $U$), the curve becomes
\begin{equation}
f(z_0,z_1,z_2,z_3)+{<v>\over \mu^{\sigma}} z^\alpha=0
\end{equation}
with $0<\sigma<1$ whose value is determined by the monomial $z^\alpha$. We would like to know the new SCFT at the deep IR, which could be 
achieved by a scaling limit as we have done in \cite{Xie:2013jc}.  

Here we give a much easier method to determine the IR SCFT. Let's first define a semi-quasihomogeneous isolated singularity $f$ as the following type of polynomial
\begin{equation}
f=f_1+f_2;
\end{equation}
Here $f_1$ defines an quasi-homogeneous isolated singularity , and $f_2$ consists of monomials with weights bigger than one. $f_0$ is called quasi-homogeneous piece of $f$. Now let's start 
with a quasi-homogeneous singularity $f_0$, and deform it using relevant deformations, and we have
\begin{equation}
f_{def}=f_0+f_1.
\end{equation}
If $f_{def}$ is a semi-quasihomogenous singularity,  $f_{def}$ can be written as $f_{def}=f_1^{'}+f_2^{'}$ and the 
IR theory is described by the quasi-homogenous piece $f_{1}^{'}$. Using this method, it is easy to determine the singularity corresponding to the IR theory by turning on 
relevant deformation $m$.

More generally we could also form a deformation by turning on deformations involving more than one terms, this 
means that we also turn on expectation value for operators with scaling dimension bigger than one. For example, let's consider the $A_2$ singularity and the following deformation
\begin{align}
&F(z,t)=z_0^2+z_1^2+z_2^2+z_3^3+3tz_3^2\rightarrow \nonumber\\
&F(z,t)=z_0^2+z_1^2+z_2^2+z_3^{'3}-3t^2z_3^{'}+2t^3,~~~z_3^{'}=z_3+t.
\end{align}
Notice that  the deformation in first line is not a form of mini-versal deformation, while the second one is written in the mini-versal deformation form by changing the coordinates. Physically, 
it means that we turn on the relevant and the coulomb branch expectation value simultaneously. 
We can take a scaling limit such that $z_3^3$ term becomes irrelevant, and the new singularity is just the $A_1$ singularity.  Alternatively, since the new singularity is semi-quasihomogenous, and 
the quasi-homogeneous piece is just $A_1$ singularity.

\begin{figure}
    \centering
    \includegraphics[width=5.0in]{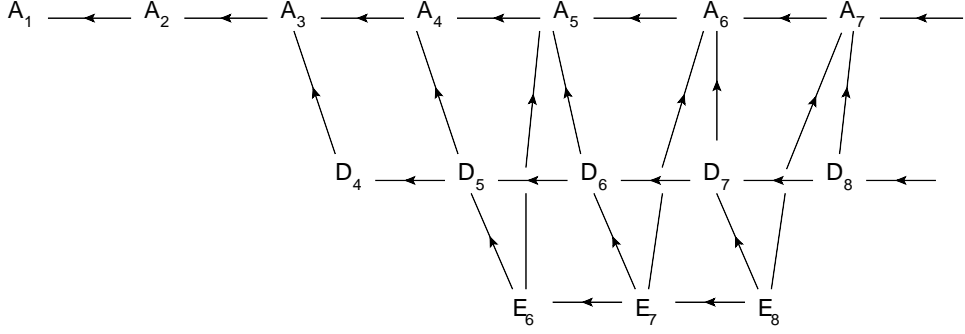}
    \caption{Adjacency of singularities which is interpreted as a RG web.}
    \label{adj}
\end{figure}

Using the above type of RG flow, one can connect the singularities, in particular, we would like to find two theories whose charge lattice 
differ by dimension one and they can be connected by a RG flow. This web is the adjacencies of singularities studied by Arnold et al \cite{arnold2012singularities}, see
figure. \ref{adj} for an example.

\subsection{Summary}
Let's summarize how to compute various physical quantities of $\mathcal{N}=2$ SCFT from singularity theory:
\begin{itemize}
\item The SW geometry is given by the mini-versal deformation of the singularity:
\begin{equation}
F(z,\lambda)=f(z)+\sum_{i=1}^\mu \lambda_\alpha\phi_\alpha(z)=0,
\end{equation}
here $\phi_i$ is the basis for the Jacobi algebra $J(f)=\mathbb{C}[z_0,z_1,z_2,z_3]/({\partial f\over z_0},{\partial f\over z_1},{\partial f\over z_2},{\partial f\over z_3})$. 
The scaling dimension of $\lambda_\alpha$ is given by the formula:
\begin{equation}
[\lambda_\alpha]={1-Q_\alpha\over \sum_{i=1}^3 q_i-1}.
\end{equation}
\item The SW differential is $\Omega={dz_0\wedge dz_1\wedge dz_2\wedge dz_3 \over dF}$ and the low energy effective action is encoded in period integral 
\begin{equation}
\lambda\rightarrow (\int_{\delta_1(\lambda)}\Omega, \ldots, \int_{\delta_\mu(\lambda)}\Omega);
\end{equation}

\item The monodromy around a co-dimensional one singularity is given by the Picard-Lefchetz transformation, 
\begin{equation}
T:x\rightarrow x+<x,\Delta_i>\Delta_i.
\end{equation}

\item The intersection form on the distinguished basis associated with vanishing cycle is conjectured to be the BPS quiver.

\item The central charge $a,c$ can be found using the formula 
\begin{align}
 & R(A)=\sum_{[u_i]>1}([u_i]-1),~~R(B)={\mu\over 2(2-\hat{c})}={\mu\over 4(\sum q_i-1)};\nonumber\\
 & a={R(A)\over 4}+{R(B)\over6}+{5 r\over 24},~~c={R(B)\over 3}+{r\over 6}.
\end{align}
\end{itemize}

The Milnor number $\mu$ of the singularity plays a crucial role for the physical property of $\mathcal{N}=2$ theory. Here let's summarize its various meanings:
\begin{itemize}
\item It is the dimension of the Jacobi algebra and the dimension of the base of mini-versal deformation of the singularity. Physically, it is equal to the number of deformations (those 
with negative scaling dimensions are irrelevant deformations) in $\mathcal{N}=2$ geometry;
\item It is equal to the dimension of middle cohomology of Milnor fibration. Physically, it is equal to the dimension of the charge lattice.
\item It is equal to the number of co-dimension one $A_1$ singularity in the mini-versal deformations. Physically, this is equal to the number of co-dimensional one singularity at which there is 
an extra massless hypermultiplet.  
\end{itemize}

\section{Classification of  hypersurface singularity}
As discussed in last section, the isolated hypersurface singularity $f(z_0,z_1,z_2,z_3)$ which seems to define a $\mathcal{N}=2$ SCFT has to satisfy the following conditions: 
\begin{itemize}
\item It has a good $\mathbb{C}^*$ action, namely 
\begin{equation}
f(\lambda^{q_i}z_i)=\lambda f_i(z),~~q_i>0;
\end{equation}
\item The weights satisfy the following condition: $\sum q_i>1$;
\end{itemize}
We would like to classify all such hypersurface singularity. Interestingly, same type of singularities have been studied by Yau and Yu \cite{yau2003classification}, and 
they give a full classification. Mathematically, what they are interested are rational isolated hypersurface singularity with a $\mathbb{C}^*$ action. The rational 
condition for the hypersurface is nothing but the condition two listed above. 
In the following, we simply take their results and reorganize them appropriately. 

To start with, we first classify isolated hypersurface singularity $f$ with a $\mathbb{C}^*$ action. It is easy to see that in order that there is an isolated 
singularity at $z_0=0$,  $f$ has to include one of the following terms $(z_0^k,z_1 z_0^k, z_2 z_0^k, z_3z_0^k)$. Similar condition is applied to 
other three variables. So we need to select one monomial within four possibilities for one variable, and there are a total of 256 possibilities. After
eliminating the simple equivalence class after permuting the variables, the polynomials  are classified into 19 types, see table. \ref{19type}. 
The weights $q_i$ of them are listed in table. \ref{19type}.  Once the weights are given, the Milnor number $\mu$ is easy to compute, and it is equal to 
\begin{equation}
\mu=\prod({1\over q_i}-1)
\end{equation}
See table. \ref{19milnor}. This classification solves the first condition. In the following subsections, we are going to impose the condition $\sum q_i>1$ for each type and list
all possible solutions. 

Not all the solutions in different types are distinct. An important criteria for the equivalence of two isolated hypersurface singularity is that 
their Jacobi algebra is isomorphic as a graded algebra \cite{mather1982classification}. Since the basis of Jacobi algebra determines the $\mathcal{N}=2$ geometry and 
the spectrum on the Coulomb branch, this indicates that the Coulomb branch spectrum determines the theory which can be defined by 
hypersurface singularity. Since the Poincare polynomial for the Jacobi algebra of   IHS with a $\mathbb{C}^*$ action is determined only by the weights $w_i$ and degree $d$, 
if we find the weights and degree  of two different polynomials $f$ and $f^{'}$ are the same, we know that they are isomorphic. 
In practice,  one can also find simple isomorphism between $f$ and $f^{'}$ if $f-f^{'}$ are sum of  weight one monomials. 

One can represent the singularity by Newton polytope in $R^4$. Let's assume that the singularity takes the following form
\begin{equation}
f=\sum a_\alpha z^{\alpha}
\end{equation}
The lattice points of the Newton polytope are $\alpha$s with $a_\alpha\neq 0$. For the hypersurface singularity with a $\mathbb{C}^*$ action, the 
points are on the hyperplane defined by the following equation
\begin{equation}
{x_0  q_0}+{x_1 q_1}+{x_2 q_2}+{x_3 q_3}=1.
\end{equation}
The Newton polytope is formed by the points $(a_\alpha+b), ~a_\alpha\neq0,~~b\in R^{4+}$ . The condition $\sum q_i>1$ implies that there are
is no points $(n_0, n_1, n_2, n_3)$ outside Newton polytope such that $(n_0+1,n_1+1,n_2+1,n_3+1)$ is in the interior of Newton polytope.

Once the normal form and Milnor number of a IHS is given, one can easily compute various interesting physical quantities:
\begin{itemize}
\item The monomial basis of Jacobi algebra can be computed using Singular \cite{DGPS}, and $\mathcal{N}=2$ geometry is given by 
\begin{equation}
F(z,\lambda)=f(z)+ \sum_{i=1}^\mu \lambda_i\phi_i(z),
\end{equation}
and one can also find the scaling dimensions of $\lambda_i$ pretty easily. Once the spectrum is known, one can list the number of 
Coulomb branch operators, the number of mass parameters, and the number of exact marginal deformations, etc. This is helpful 
for the classification purpose.
\item The central charge $a$ and $c$ can be extracted from the information of spectrum and the Milnor number. 
\end{itemize}
The above computations can be made using the computer software Singular \cite{DGPS}.

\begin{table}
\begin{center}
  \begin{tabular}{|l|c|c| }
    \hline
    Type & $f(z_0,z_1,z_2,z_3)$ & $\sum q_i$  \\ \hline
    I & $z_0^a+z_1^b+z_2^c+z_3^d$& ${1\over a}+ {1\over b}+{1\over c}+{1\over d}$\\ \hline
       II & $z_0^a+z_1^b+z_2^c+z_2z_3^d$& ${1\over a}+ {1\over b}+{1\over c}+{c-1\over cd}$\\ \hline
       III & $z_0^a+z_1^b+z_2^cz_3+z_2z_3^d$& ${1\over a}+ {1\over b}+{d-1\over cd-1}+{c-1\over cd-1}$\\ \hline
       IV & $z_0^a+z_0z_1^b+z_2^c+z_2z_3^d$& ${1\over a}+ {a-1\over a b}+{1\over c}+{c-1\over cd}$\\ \hline
       V & $z_0^az_1+z_0z_1^b+z_2^c+z_2z_3^d$& ${b-1\over ab-1}+ {a-1\over a b-1}+{1\over c}+{c-1\over cd}$\\ \hline
       VI & $z_0^az_1+z_0z_1^b+z_2^cz_3+z_2z_3^d$& ${b-1\over ab-1}+ {a-1\over a b-1}+{d-1\over c}+{cd-1\over cd}$\\ \hline
       VII & $z_0^a+z_1^b+z_1z_2^c+z_2z_3^d$& ${1\over a}+ {1\over b}+{b-1\over b c}+{b(c-1)+1\over bcd}$\\ \hline
       VIII & $z_0^a+z_1^b+z_1z_2^c+z_1z_3^d+z_2^pz_3^q$,  & ${1\over a}+ {1\over b}+{b-1\over b c}+{b-1\over bd}$\\ 
 ~ &  ${p(b-1)\over bc}+{q(b-1)\over bd}=1$& ~\\  \hline
       IX & $z_0^a+z_1^bz_3+z_2^cz_3+z_1z_3^d+z_1^pz_2^q$,  & ${1\over a}+ {d-1\over bd-1}+{b(d-1)\over c(bd-1)}+{b-1\over bd-1}$\\ 
 ~ &  ${p(d-1)\over bd-1}+{qb(d-1)\over c(bd-1)}=1$& ~\\  \hline
        X & $z_0^a+z_1^bz_2+z_2^cz_3+z_1z_3^d$& ${1\over a}+ {d(c-1)+1\over bcd+1}+{b(d-1)+1\over bcd+1}+{c(b-1)+1\over bcd+1}$\\ \hline
                XI & $z_0^a+z_0z_1^b+z_1z_2^c+z_2z_3^d$& ${1\over a}+ {a-1\over ab}+{a(b-1)+1\over abc}+{ab(c-1)+(a-1)\over abcd}$\\ \hline

                XII & $z_0^a+z_0z_1^b+z_0z_2^c+z_1z_3^d+z_1^pz_2^q$& ${1\over a}+ {a-1\over ab}+{a-1\over ac}+{a(b-1)+1\over abd}$\\ 
                ~&${p(a-1)\over ab}+{q(a-1)\over ac}=1$&~ \\ \hline
  XIII & $z_0^a+z_0z_1^b+z_1z_2^c+z_1z_3^d+z_2^pz_3^q$& ${1\over a}+ {a-1\over ab}+{a-1\over ac}+{a(b-1)+1\over abd}$\\ 
                ~&${p(a(b-1)+1)\over abc}+{q(a(b-1)+1)\over abd}=1$&~ \\ \hline
                  XIV & $z_0^a+z_0z_1^b+z_0z_2^c+z_0z_3^d+z_1^pz_2^q+z_2^rz_3^s$& ${1\over a}+ {a-1\over ab}+{a-1\over ac}+{a-1\over ad}$\\ 
                ~&${p(a-1)\over ab}+{q(a-1)\over ac}=1={r(a-1)\over ac}+{s(a-1)\over ad}$&~ \\ \hline
                XV&$z_0^az_1+z_0z_1^b+z_0z_2^c+z_2z_3^d+z_1^pz_2^q$& ${b-1\over ab-1}+{a-1\over ab-1}+{b(a-1)\over c(ab-1)}+{c(ab-1)-b(a-1)\over cd(ab-1)}$ \\ 
                ~&${p(a-1)\over ab-1}+{qb(a-1)\over c(ab-1)}=1$&~ \\ \hline
                 XVI&$z_0^az_1+z_0z_1^b+z_0z_2^c+z_0z_3^d+z_1^pz_2^q+z_2^rz_3^s$& ${b-1\over ab-1}+{a-1\over ab-1}+{b(a-1)\over c(ab-1)}+{b(a-1)\over d(ab-1)}$ \\ 
                ~&${p(a-1)\over ab-1}+{qb(a-1)\over c(ab-1)}=1={r(a-1)\over ac}+{s(a-1)\over ad}$&~ \\ \hline
                    XVII&$z_0^az_1+z_0z_1^b+z_1z_2^c+z_0z_3^d+z_1^pz_2^q+z_0^rz_2^s$& ${b-1\over ab-1}+{a-1\over ab-1}+{a(b-1)\over c(ab-1)}+{b(a-1)\over d(ab-1)}$ \\ 
                ~&${p(a-1)\over ab-1}+{qb(a-1)\over d(ab-1)}=1={r(b-1)\over ab-1}+{sa(b-1)\over c(ab-1)}$&~ \\ \hline
                XVIII&$z_0^az_2+z_0z_1^b+z_1z_2^c+z_1z_3^d+z_2^pz_3^q$&${b(c-1)+1\over abc+1}+{c(a-1)+1\over abc+1}+{a(b-1)+1\over c(abc+1)}+{c(a(b-1)+1)\over d(abc+1)}$ \\
                ~&${p(a(b-1)+1)\over abc+1}+{qc[a(b-1)+1]\over d(abc+1)}=1$&~ \\ \hline
                XIX &$z_0^a+z_0z_1^b+z_2^cz_1+z_2z_3^d$ &${[b(d(c-1)+1)-1\over abcd-1}+{[d(c(a-1)+1)-1\over abcd-1}$\\
                ~&~&$+{[a(b(d-1)+1)-1\over abcd-1}+{[c(a(b-1)+1)-1\over abcd-1}$ \\
                
    \hline
  \end{tabular}
      \caption{The canonical form for isolated hypersurface singularity with a good $\mathbb{C}^*$ action  whose weights on coordinates $z_i$ are also given.}
      \label{19type}
\end{center}
\end{table}

\begin{table}[h]
\begin{center}
  \begin{tabular}{|l|c|c| }
    \hline
    Type & $f(z_0,z_1,z_2,z_3)$ & $\mu$  \\ \hline
    I & $z_0^a+z_1^b+z_2^c+z_3^d$& $\mu=(a-1)(b-1)(c-1)(d-1)$\\ \hline
       II & $z_0^a+z_1^b+z_2^c+z_2z_3^d$& $\mu=(a-1)(b-1)[c(d-1)+1]$\\ \hline
       III & $z_0^a+z_1^b+z_2^cz_3+z_2z_3^d$& $\mu=(a-1)(b-1)cd$\\ \hline
       IV & $z_0^a+z_0z_1^b+z_2^c+z_2z_3^d$& $\mu=[a(b-1)+1][c(d-1)+1]$\\ \hline
       V & $z_0^az_1+z_0z_1^b+z_2^c+z_2z_3^d$& $\mu=ab[c(d-1)+1]$\\ \hline
       VI & $z_0^az_1+z_0z_1^b+z_2^cz_3+z_2z_3^d$& $\mu=abcd$\\ \hline
       VII & $z_0^a+z_1^b+z_1z_2^c+z_2z_3^d$& $\mu=(a-1)[bc(d-1)+b-1]$\\ \hline
       VIII & $z_0^a+z_1^b+z_1z_2^c+z_1z_3^d+z_2^pz_3^q$,  & $\mu={(a-1)[b(c-1)+1][b(d-1)+1]\over b-1}$\\ 
 ~ &  ${p(b-1)\over bc}+{q(b-1)\over bd}=1$& ~\\  \hline
       IX & $z_0^a+z_1^bz_3+z_2^cz_3+z_1z_3^d+z_1^pz_2^q$,  & $\mu={(a-1)d[c(bd-1)-b(d-1)]\over d-1}$\\ 
 ~ &  ${p(d-1)\over bd-1}+{qb(d-1)\over c(bd-1)}=1$& ~\\  \hline
        X & $z_0^a+z_1^bz_2+z_2^cz_3+z_1z_3^d$& $\mu=(a-1)bcd$\\ \hline
                XI & $z_0^a+z_0z_1^b+z_1z_2^c+z_2z_3^d$& $\mu=abc(d-1)+a(b-1)+1$\\ \hline

                XII & $z_0^a+z_0z_1^b+z_0z_2^c+z_1z_3^d+z_1^pz_2^q$& $\mu={(a(c-1)+1)(ab(d-1)+a-1)\over a-1}$\\ 
                ~&${p(a-1)\over ab}+{q(a-1)\over ac}=1$&~ \\ \hline
  XIII & $z_0^a+z_0z_1^b+z_1z_2^c+z_1z_3^d+z_2^pz_3^q$& $\mu={[ab(c-1)+a-1][ab(d-1)+a-1]\over a(b-1)+1}$\\ 
                ~&${p(a(b-1)+1)\over abc}+{q(a(b-1)+1)\over abd}=1$&~ \\ \hline
                  XIV & $z_0^a+z_0z_1^b+z_0z_2^c+z_0z_3^d+z_1^pz_2^q+z_2^rz_3^s$& $\mu={[a(b-1)+1][a(c-1)+1][a(d-1)+1]\over (a-1)^2}$\\ 
                ~&${p(a-1)\over ab}+{q(a-1)\over ac}=1={r(a-1)\over ac}+{s(a-1)\over ad}$&~ \\ \hline
                XV&$z_0^az_1+z_0z_1^b+z_0z_2^c+z_2z_3^d+z_1^pz_2^q$& $\mu={a[c(d-1)(ab-1)+b(a-1)]\over a-1}$ \\ 
                ~&${p(a-1)\over ab-1}+{qb(a-1)\over c(ab-1)}=1$&~ \\ \hline
                 XVI&$z_0^az_1+z_0z_1^b+z_0z_2^c+z_0z_3^d+z_1^pz_2^q+z_2^rz_3^s$& $\mu={a[c(ab-1)-b(a-1)][d(ab-1)-b(a-1)]\over b(a-1)^2}$ \\ 
                ~&${p(a-1)\over ab-1}+{qb(a-1)\over c(ab-1)}=1={r(a-1)\over ac}+{s(a-1)\over ad}$&~ \\ \hline
                    XVII&$z_0^az_1+z_0z_1^b+z_1z_2^c+z_0z_3^d+z_1^pz_2^q+z_0^rz_2^s$& $\mu={[c(ab-1)-a(b-1)][d(ab-1)-b(a-1)]\over (a-1)(b-1)}$ \\ 
                ~&${p(a-1)\over ab-1}+{qb(a-1)\over d(ab-1)}=1={r(b-1)\over ab-1}+{sa(b-1)\over c(ab-1)}$&~ \\ \hline
                XVIII&$z_0^az_2+z_0z_1^b+z_1z_2^c+z_1z_3^d+z_2^pz_3^q$&$\mu={ab[abc(d-1)+c(a-1)+d]\over a(b-1)+1}$ \\
                ~&${p(a(b-1)+1)\over abc+1}+{qc[a(b-1)+1]\over d(abc+1)}=1$&~ \\ \hline
                XIX &$z_0^a+z_0z_1^b+z_2^cz_1+z_2z_3^d$ &~\\
                ~&~&$\mu=abcd$ \\
                \hline
   \end{tabular}
  \caption{The Milnor number of the Hypersurface singularities.}
  \label{19milnor}
 \end{center}
 \end{table}

\newpage 

\subsection{Type I}
Here $f=z_0^a+z_1^b+z_2^c+z_3^d$, and $a,b,c,d\geq 2$ so that there is isolated singularity at the origin. The $\sum q_i>1$ condition is 
\begin{equation}
{1\over a}+{1\over b}+ {1\over c}+{1\over d}>1.
\label{typeI}
\end{equation}
There is an obvious symmetry exchanging $(a,b,c,d)$, so we can require $a\leq b\leq c\leq d$. 

The solutions to inequality $\ref{typeI}$ are separated in seven infinite sequences which are actually already studied (we also list their names used in literature)
\begin{table}[h]
\begin{center}
  \begin{tabular}{|c|c| }
    \hline
Solution& Other name \\ \hline
 (2,2,p,q)&$(A_{p-1}, A_{q-1})$ \\ \hline 
(2,3,3,k)&$(D_4, A_{k-1})$ \\ \hline
(2,3,4,k)&$(E_6, A_{k-1})$ \\ \hline
(2,3,5,k)& $(E_8,A_{k-1})$  \\ \hline
(2,3,6,k)& $(E_8^{1,1}, A_{k-1})$ \\ \hline
(2,4,4,k)& $(E_7^{1,1},A_{k-1} )$ \\ \hline
(3,3,3,k)& $(E_6^{1,1},A_{k-1})$  \\ 
    \hline
  \end{tabular}
  \caption{Infinite sequence of type I singularity.}
  \label{Iinf}
\end{center}
\end{table}
The first four classes have been studied in \cite{Cecotti:2010fi}, and they 
can be also engineered using M5 brane with the $ADE$ type on a sphere with a single irregular singularity \cite{Xie:2012hs,Wang:2015mra}.
The last three classes always have exact marginal deformations and can be described by a weakly coupled gauge theories, and some aspects of these theories are studied in \cite{DelZotto:2015rca}.
There are also 13 class of sporadic examples, see table. \ref{Ifin}.
\begin{table}[h]
\begin{center}
  \begin{tabular}{|c|c||c|c||c|c| }
    \hline
  (2,3,7,k)&$6<k<42$&(2,3,8,k)&$7<k<24$&(2,3,9,k)&$8<k<18$ \\ \hline 
    (2,3,10,k)&$9<k<15$&(2,3,11,k)&$10<k<14$&(2,4,5,k)&$4<k<20$ \\ \hline 
    (2,4,6,k)&$5<k<12$&(2,4,7,k)&$6<k<10$&(2,5,5,k)&$4<k<10$ \\ \hline 
    (2,5,6,k)&$5<k<8$&(3,3,4,k)&$3<k<12$&(3,3,5,k)&$4<k<8$ \\  \hline
    (3,4,4,k)&$3<k<6$&~&~&~&~ \\
    \hline
  \end{tabular}
  \caption{Sporadic sequence of type I singularity.}
  \label{Ifin}
\end{center}
\end{table}

\subsection{Type II}
Here $f=z_0^a+z_1^b+z_2^c+z_2z_3^d$, and we require $a,b,c\geq 2,~d\geq 1$ to have an isolated singularity at the origin. There are some overlap with Type I singularity. 
Since the weights of $z_3$ is ${c-1\over d c}$, if we can find 
integer $n$ such that ${n (c-1)\over d c}=1$, then $z_3^n$ is an exact marginal deformation and the singularity can be put in the equivalent form $f^{'}=z_0^a+z_1^b+z_2^c+z_3^n$,
which is a type I singularity. It is easy to find out that the reducible condition is
\begin{equation}
{d\over c-1}\in \mathbb{Z}.
\label{equi2}
\end{equation}
The $\sum q_i>1$ inequality is 
\begin{equation}
{1\over a}+ {1\over b}+{1\over c}+{c-1\over cd}>1,
\label{type II}
\end{equation}
This equation is symmetric in exchanging $(a,b)$ and $(c,d)$, so we could require $a\leq b$ and $c\leq d$ in solving the inequality.  
The exchange of $a,b$ gives us equivalent singularity, while the exchange of $(c,d)$ gives us different singularity.  

We list the solutions by requiring $a\leq b$ and $c\leq d$ and one should keep in mind that exchanging $c$ and $d$ would give 
us inequivalent theory. Some of the solutions can be reduced to type I using the reduction discussed above. More generally, if 
we find the same spectrum from the deformations with certain type I singularity, then those two singularities give the same 4d SCFT.

The infinite sequences of solutions are listed in table. \ref{IIinf}, and the sporadic sequences are listed in table. \ref{IIfin}. 
\begin{table}
\begin{center}
  \begin{tabular}{|c|c|c|c|c| }
    \hline
 (r,s,t,1)&(2,2,r,s)&$(2,r,2,s)$&(3,s,2,2)&(4,s,2,2) \\ \hline 
 (2,3,3,s)&(2,3,4,s)&((2,3,5,s)&(2,3,6,s)&(2,4,3,s) \\ \hline
 (2,5,3,s)&(2,6,3,s)&(2,s,3,3)&(2,s,3,4)&(3,3,2,s) \\ \hline
 (3,4,2,s)&(3,5,2,s)&(3,6,2,s)&(3,s,2,3)&(2,4,4,s) \\ \hline
 (4,4,2,s)&(3,3,3,s)&&&\\ \hline
  \end{tabular}
  \caption{Infinite sequences of type II singularity.}
  \label{IIinf}
\end{center}
\end{table}

\begin{table}
\begin{center}
  \begin{tabular}{|c|c||c|c||c|c| }
    \hline
    (5,k,2,2)&$4<k<20$&(6,k,2,2)&$5<k<12$&(7,k,2,2)&$6<k<10$ \\ \hline
  (2,3,7,k)&$6<k<36$&(2,3,8,k)&$7<k<21$&(2,3,9,k)&$8<k<16$ \\ \hline 
  (2,3,10,k)&$9<k<14$&(2,3,11,k)&$10<k<12$& (2,7,3,k)&$6< k<28$ \\ \hline
   (2,8,3,k)& $7<k<16$&  (2,9,3,k)& $8<k<12$&  $(2,k,3,5)$& $5<k<30$   \\ \hline
$(2,k,3,6)$&$6<k<18$&$(2,k,3,7)$&$7<k<14$& $(2,k,3,8)$&$8<k<12$\\ \hline 
$(2,k,3,9)$&$9<k<11$&(3,7,2,k)&$6<k<21$&(3,8,2,k)&$7<k<12$ \\ \hline
(3,k,2,4)& $4<k<24$& (3,k,2,5)&$5<k<15$ &(3,k,2,6)& $6<k<12$ \\ \hline
(3,k,2,8)& $8<k<10$ &(4,k,2,3)& $3<k<12$ &(5,k,2,3) & $4<k<8$ \\ \hline
(2,4,5,k)&$4<k<16$&(2,4,6,k)&$5<k<10$&
 $(2,4,7,k)$&  $6<k<8$ \\ \hline
 (2,6,4,k)&$ 5<k<9$&(2,k,4,4)&$4<k<16$&(2,k,4,5)& $ 5<k<10$ \\ \hline
 (2,k,4,6) &$ 6<k<8$& (4,5,2,k)&$4<k<10$&(4,k,2,4)&$4<k<8$ \\ \hline
(4,k,2,5)&$5<k<7$&(2,5,5, k)&$4<k<8$ &(2,5,6,k)& $ 5<k<7$ \\ \hline
 (2,k,5,5)& $ 5<k<8$ &(3,3,4,k)&$3<k<9$ &(3,3,5,k)& $4<k<6$    \\ \hline
 (3,4,3,k)& $ 3<k<8$ &(3,k,3,3)& $ 3<k<9$ & (3,k,3,4)& $4<k<6$ \\ \hline
 (4,k,3,3)& $ 3<k<6$ &(3,4,4,4)&(5,5,2,4)& (2,5,4,k)&  $4<k<16$ \\ 
    \hline
  \end{tabular}
  \caption{Sporadic sequences of type II singularity.}
  \label{IIfin}
\end{center}
\end{table}

\subsection{Type III}
Here $f=z_0^a+z_1^b+z_2^cz_3+z_2z_3^d$, and we require $a,b,c,d\geq 2$ so that there is only an isolated singularity at the origin. There is an obvious symmetry exchanging $a,b$, and $c,d$, so we 
require $b\geq a$ and $d\geq c$. The weights of $z_2$ and $z_3$ are 
\begin{equation}
q(z_2)={d-1\over cd -1},~~q(z_3)= {c-1\over cd -1}
\end{equation}
If we can find an integer $n$ such that $n q(z_2) ~\text{or}~ nq(z_3)=1$, we can reduce it to type II singularity. The solutions can be easily found, i.e
\begin{equation}
{c-1\over d-1}~\text{or}~{d-1\over c-1} \in \mathbb{Z} 
\end{equation}

The inequality we are going to solve is 
\begin{equation}
{1\over a}+ {1\over b}+{d-1\over cd-1}+{c-1\over cd-1}>1.
\end{equation}
The  infinite sequences of solutions are listed in table. \ref{IIIinf}, and the sporadic sequences are listed in table . \ref{IIIfin} 
\begin{table}
\begin{center}
  \begin{tabular}{|c|c|c|c|c| }
    \hline
 (2,2,r,s)&(2,r,2,s)&$(3,s,2,2)$&(2,3,3,s)&(2,3,4,s) \\ \hline 
 (2,3,5,s)&(2,3,6,s)&(2,4,3,s)&(2,5,3,s)&(2,6,3,s)\\ \hline
 (2,s,3,3)&(3,3,2,s)&(3,4,2,s)&(3,5,2,s)&(3,6,2,s) \\ \hline
 (2,4,4,s)&(4,4,2,s)&(3,3,3,s)&~&~\\ \hline
  \end{tabular}
  \caption{Infinite sequence of type III singularity.}
  \label{IIIinf}
\end{center}
\end{table}
\begin{table}
\begin{center}
  \begin{tabular}{|c|c||c|c||c|c| }
    \hline
    (4,k,2,2)&$3<k<12$&(5,k,2,2)&$4<k<8$& (2,3,7,k)&$6<k<31$ \\ \hline
 (2,3,8,k)&$7<k<19$&(2,3,9,k)&$8<k<15$ & (2,3,10,k)&$9<k<13$ \\ \hline
 (2,7,3,k)&$6<k<19$&(2,8,3,k)&$7<k<11$ &(2,k,3,4)&$4<k<22$  \\ \hline
  (2,k,3,5)&$5<k<14$& (2,k,3,6)&$6<k<12$ &(2,k,3,7)&$7<k<10$  \\ \hline
 (2,k,3,8)&$8<k<10$&(3,7,2,k)&$6<k<11$ &(3,k,2,3)&$3<k<15$   \\ \hline
(3,k,2,4)&$4<k<11$&(3,k,2,5)&$5<k<9$&(3,k,2,6)&$6<k<9$ \\ \hline
  (2,4,5,k)& $ 4<k<13$&(2,4,6,k)&$5<k<9$&(2,5,4,k)&$ 4<k<12$ \\ \hline
 (2,6,4,k)& $   5<k<7$ & (2,k,4,4)&$   4<k<10$&(2,k,4,5)& $ 5<k<8$ \\ \hline
 (2,5,5,k)& $ 4<k<7$ &(3,3,4,k)&$ 3<k<7$ &(3,4,3,k)& $3<k<6$ \\ \hline
 (3,k,3,3)& $ 3<k<6$&(4,5,2,5)&(4,5,2,4)&~&~ \\ 
    \hline
  \end{tabular}
   \caption{Finite sequence of type III singularity.}
  \label{IIIfin}
\end{center}
\end{table}
Here again we list the possible theories and it is easy to identify them with some of type II theories. 

\subsection{Type IV}
Here $f=z_0^a+z_0z_1^b+z_2^c+z_2z_3^d$, and we have $a, c \geq 2$ and $ b,d\geq 1$ so
that there is an isolated singularity at the origin. There is an symmetry exchanging the pair $(a,b)$ and $(c,d)$, so 
we also require $d\geq b$. 
Again his can be reduced to the first two types if 
\begin{equation}
{b-1\over a}\in \mathbb{Z}~~\text{or}~~{d-1\over c}\in \mathbb{Z}
\end{equation}

The inequality we want to solve is 
\begin{equation}
{1\over a}+ {a-1\over a b}+{1\over c}+{c-1\over cd}>1,
\end{equation}
For this inequality, there is an symmetry exchanging $(a, b)$ and $(c,d)$, so we require $b\geq a$ and $d\geq c$ 
for the solutions. One should keep in mind that the exchange of $(a,b)$ or $(c,d)$ give us inequivalent 
singularity!
The infinite sequence of solutions are listed in table .\ref{IVinf}, and the sporadic sequence of solutions are 
listed in table. \ref{IVfin}.
\begin{table}[h]
\begin{center}
  \begin{tabular}{|c|c|c|c|c| }
    \hline
 (r,1,s,t)&(r,1,s,1)&$(2,2,2,s)$&(2,2,3,s)&(2,2,4,s) \\ \hline 
 (2,r,2,s)&(2,3,3,s)&(2,s,3,3&(2,s,3,4)& \\ \hline
  \end{tabular}
 \caption{Infinite sequence of type IV singularity.}
  \label{IVinf}
\end{center}
\end{table}
\begin{table}[h]
\begin{center}
  \begin{tabular}{|c|c||c|c||c|c| }
    \hline
  (2,2,5,k)&$4<k<16$&(2,2,6,k)&$5<k<10$&(2,2,7,k)&$6<k<8$ \\ \hline 
  (2,3,4,k)&$3<k<9$&(2,3,5,k)& $4<k<6$&(2,4,3,k)& $3<k<16$ \\ \hline
  (2,5,3,k)&$5<k<15$&(2,6,3,k)&$ 6<k<9$& (2,4,4,k)& $3<k<7$ \\ \hline
  (2,k,4,4)& $4<k<8$ &(3,3,3,k)& $2<k<6$&&  \\ \hline
  
  \end{tabular}
   \caption{Sporadic sequence of type IV singularity.}
  \label{IVfin}
\end{center}
\end{table}

\newpage
\subsection{Type V}
The polynomial is $f=z_0^az_1+z_0z_1^b+z_2^c+z_2z_3^d$, we have $a, b, c\geq 2,~d\geq 1$ so
that there is an isolated singularity at the origin. There is a symmetry exchanging the pair $(a,b)$.
The singularity is reduced to previous type if 
\begin{equation}
{b-1 \over a-1}\in \mathbb{Z}~\text{or}~{d\over c-1}\in \mathbb{Z}
\end{equation}
The inequality we want to solve is 
\begin{equation}
{b-1\over ab-1}+ {a-1\over a b-1}+{1\over c}+{c-1\over cd}>1,
\end{equation}
and this inequality is invariant under exchange of $c$ and $d$, so 
the solution is listed by requiring $d\geq c$, however the singularity 
associated with these two different ordering are different!
The infinite sequence of solutions are listed in table. \ref{Vinf}, and 
the sporadic sequence of solutions are listed in table. \ref{Vfin}.

\begin{table}[h]
\begin{center}
  \begin{tabular}{|c|c|c|c|c| }
    \hline
 (r,s,t,1)&(2,2,2,s)&((2,2,3,s)&(2,r,2,s)&(3,s,2,2) \\ \hline 
 (4,s,2,2)&(2,s,3,3)&(2,s,3,4)&(3,3,2,s)&(3,s,2,3)\\ \hline
  \end{tabular}
  \caption{Infinite sequence of type V singularity.}
  \label{Vinf}
\end{center}
\end{table}
\begin{table}[h]
\begin{center}
  \begin{tabular}{|c|c||c|c||c|c| }
    \hline
  (2,2,4,k)&$3<k<9$&(2,2,5,k)&$4<k<6$ &(5,k,2,2)&$4<k<13$ \\ \hline 
  (5,k,2,2)& $4<k<13$ &(6,k,2,2)& $5<k<9$ & (2,3,3,k)&$2<k<10$ \\ \hline
  (2,3,4,4)& $3<k<5$ &(2,4,3,k)& $3<k<7$ & (2,5,3,k)& $4<k<6$ \\ \hline
  (2,k,3,5)&$5<k<8$&(3,4,2,k)& $4<k<11$ &(3,5,2,k)& $5<k<8$ \\ \hline
  (4,k,2,3)& $3<k<7$ &(2,4,4,4)&(4,4,2,4)&(3,3,3,3)&(3,4,3,3) \\ \hline

  \end{tabular}
   \caption{Sporadic sequence of type V singularity.}
  \label{Vfin}
\end{center}
\end{table}

\subsection{Type VI}
The polynomial is $f=z_0^az_1+z_0z_1^b+z_2^cz_3+z_2z_3^d$, we have $a, b, c,d\geq 2$ so
that there is an isolated singularity at the origin. Due to symmetry, we require $b\geq a$, $d\geq c$, and 
$c\geq a$. The singularity is reduced to previous type if 
\begin{equation}
{b-1 \over a-1}\in \mathbb{Z}~\text{or}~{d-1\over c-1}\in \mathbb{Z}.
\end{equation}
The inequality we want to solve is 
\begin{equation}
{b-1\over ab-1}+ {a-1\over a b-1}+{d-1\over c}+{cd-1\over cd}>1
\end{equation}
The infinite sequence of solutions are listed in table. \ref{VIinf}. All of them can be reduced to previous type, so there is no new infinite sequence of solutions. 
The sporadic set of examples are listed in table. \ref{VIfin}, and all of them can also be reduced to previous type of singularities. Therefore there is 
no new solutions for this class of singularity.

\begin{table}[h]
\begin{center}
  \begin{tabular}{|c|c|c|c| }
    \hline
 (2,2,2,s)&(2,2,3,s)&(2,r,2,s)&(2,s,3,3) \\ \hline 
  \end{tabular}
  \caption{Infinite sequence of type VI singularity.}
  \label{VIinf}
\end{center}
\end{table}
\begin{table}[h]
\begin{center}
  \begin{tabular}{|c|c||c|c||c|c| }
    \hline
  (2,2,4,k)&$3<k<7$&(2,3,3,k)&$2<k<7$ &(2,4,3,4)&(2,5,3,4) \\
    \hline
  \end{tabular}
    \caption{Sporadic sequence of type VI singularity.}
  \label{VIfin}
\end{center}
\end{table}

\subsection{Type VII}
The singularity has the form $f=z_0^a+z_1^b+z_1z_2^c+z_2z_3^d$. We have $a\geq 2, b\geq 2, c\geq 1, d\geq 1$ so
that there is an isolated singularity at the origin. 
The singularity can be reduced to previous type if 
\begin{equation}
{c\over b-1}\in \mathbb{Z},~~~~\text{or}~~~~{d (b-1)\over b(c-1)+1}\in \mathbb{Z}
\end{equation}
The inequality we want to solve is 
\begin{equation}
{1\over a}+ {1\over b}+{b-1\over b c}+{b(c-1)+1\over bcd}>1.
\end{equation}
This inequality is invariant under exchange of $(b,d)$, so the solution is listed by requiring $d\geq b$. However,
two orderings give different solutions. 
The  infinite sequence of solutions are listed in table. \ref{VIIinf}, and the sporadic examples are listed in table. \ref{VIIfin}.
\begin{table}[h]
\begin{center}
  \begin{tabular}{|c|c|c|c|c| }
    \hline
 (r,s,1,t)&(r,s,t,1)&(2,r,2,s)&(2,2,r,s)&(3,2,2,s)\\ \hline 
(4,2,2,s)&(s,2,2,2)&(s,2,2,3)&(r,2,s,2)&(2,3,3,s) \\ \hline
(2,4,3,s)&(3,3,2,s)&(2,3,4,s)&(2,3,s,3)&(2,3,s,4) \\ \hline
(2,3,s,5)&(2,3,s,6)&(3,2,3,s)&(3,2,s,3)&(3,2,s,4) \\ \hline
(3,2,s,5)&(3,2,s,6)&(4,2,s,3)&(5,2,s,3)&(6,2,s,3) \\ \hline
(2,4,s,4)&(4,2,s,4)&(3,3,s,3)&& \\ \hline
  \end{tabular}
  \caption{Infinite sequence of type VII singularity.}
  \label{VIIinf}
\end{center}
\end{table}
\begin{table}[h]
\begin{center}
  \begin{tabular}{|c|c||c|c||c|c| }
    \hline
  (5,2,2,k)&$4<k<15$&(6,2,2,k)&$5<k<9$&(k,2,2,4)&$4<k<16$ \\ \hline 
  (k,2,2,5)&$5<k<10$&(k,2,2,6)&$6<k<8$&(2,5,3,k)& $4<k<22$ \\ \hline
  (2,6,3,k)& $5<k<13$ &(2,7,3,k)& $6<k<10$&(2,8,3,k)& $7<k<9$ \\ \hline
  (3,4,2,k) & $3<k<15$ & (3,5,2,k)& $4<k<9$ &(3,6,2,k)& $5<k<7$ \\ \hline
  (2,3,5,k) &$4<k<26$ &  (2,3,6,k)& $ 5<k<16$ &(2,3,7,k)& $6<k<13$ \\ \hline
  (2,3,8,k) & $7<k<11$ & (2,3,9,k) & $ 8<k<10$ & (4,3,2,k) & $3<k<8$ \\ \hline
  (2,3,k,7)& $7<k<24$ &(2,3,k,8) & $8<k<14$ &(2,3,k,9) &$1<k<11$ \\ \hline
  (k,3,2,3)& $3<k<9$ & (k,3,2,4)& $4<k<6$ &(3,2,4,k)& $3<k<21$ \\ \hline
  (3,2,5,k) & $4<k<14$ & (3,2,6,k) & $5<k<11$ & (3,2,7,k)& $6<k<10$ \\ \hline
  (3,2,8,k) & $7<k<9$ & (4,2,3,k)& $3<k<10$ & (5,2,3,k) & $4<k<7$  \\ \hline
  (3,2,k,7)& $7<k<18$ &(3,2,k,8)& $8<k<11$&(k,2,3,3)&$3<k<18$ \\ \hline
  (k,2,3,4)&$4<k<8$ &(7,2,k,3)& $6<k<14$& (k,2, 4,3)&$4<k<12$ \\ \hline
  (k,2,5,3)&$5<k<10$&(k,2,6,3)& $6<k<9$&(k,2,7,3)&$7<k<9$ \\ \hline
  (2,4,4,k)& $3<k<13$& (2,5,4,k)& $4<k<8$ &(2,6,4,k)&$5<k<7$ \\ \hline
  (4,4,2,4)&~&(2,4,5,k)&$4<k<9$&(2,4,6,k)&$5<k<7$ \\ \hline
  (2,4,k,5)&$5<k<12$&(2,4,k,6)&$6<k<8$&(4,2,4,k)&$3<k<7$ \\ \hline
  (4,2,5,k)& $4<k<6$ &(4,2,k,5)& $5<k<8$&(k,2,4,4)&$4<k<7$ \\ \hline
  (5,2,k,4)&$4<k<8$ &(2,5,5,5)&~&(2,5,6,5)&~\\ \hline
  (3,3,3,k)&$2<k<7$&(3,4,3,4)&~&(3,3,4,4)&~\\ \hline
  (3,3,5,4)&~&(k,3,3,3)&$3<k<6$&(4,3,k,3)&$3<k<6$ \\ 
    \hline
  \end{tabular}
 \caption{Sporadic sequence for type VII singularity.}
 \label{VIIfin}
\end{center}
\end{table}

\newpage
\subsection{Type VIII}
The singularity has the form $f=z_0^a+z_1^b+z_1z_2^c+z_1z_3^d+z_2^pz_3^q$ with 
 constraint ${p(b-1)\over bc}+{q(b-1)\over bd}=1$. We require $a,b\geq 2$, and $c,d\geq 1$ such
 that there is an isolated singularity at the origin. The singularity is reduced to previous class if 
\begin{equation}
{c\over b-1}\in \mathbb{Z},~~~\text{or},~~{d \over b-1}\in \mathbb{Z}
\end{equation}
Here $c$ and $d$ are totally symmetric, and we assume $d\geq c$. 
Notice that it is not always possible to find a pair of integer $(p,q)$ for a given set of integers $(a,b,c,d)$. 

The inequality we want to solve is:
\begin{equation}
{1\over a}+ {1\over b}+{b-1\over b c}+{b-1\over bd}>1,
\end{equation}
The infinite sequence of solutions are listed in table. \ref{VIIIinf}, and the sporadic examples are 
listed in table. \ref{VIIIfin}

\begin{table}[h]
\begin{center}
  \begin{tabular}{|c|c|c|c|c| }
    \hline
 (r,s,1,t)&(r,s,t,1)&(2,2,r,s)&(3,2,2,s)&(4,2,2,s)\\ \hline 
(2,r,2,s)&(s,2,2,2)&(r,s,2,2)&(2,3,3,s)&(2,3,4,s)\\ \hline
(3,2,3,s)&(2,4,3,s)&(2,s,3,3)&(2,s,3,4)&(2,s,3,5) \\ \hline
(2,s,3,6)&(3,3,2,s)&(3,s,2,3)&(3,s,2,4)&(3,s,2,5) \\ \hline
(3,s,2,6)&(4,s,2,3)&(5,s,2,3)&(6,s,2,3)&(2,s,4,4) \\ \hline
(4,s,2,4)&(3,s,3,3)&&& \\ \hline
  \end{tabular}
  \caption{Infinite sequence of type VIII singularity.}
  \label{VIIIinf}
\end{center}
\end{table}

\begin{table}[h]
\begin{center}
  \begin{tabular}{|c|c||c|c||c|c| }
    \hline
  (5,2,2,k)&$5<k<12$&(k,2,2,3)&$3<k<12$&(k,2,2,4)& $4<k<8$ \\ \hline 
 (k,2,2,5)& $5<k<7$ &(2,3,5,k)&$4<k<20$&(2,3,6,k)& $5<k<12$ \\ \hline
 (2,3,7,k)& $6<k<10$ &(3,2,4,k) &$3<k<12$ &(3,2,5,k)& $4<k<8$ \\ \hline
 (2,5,3,k)& $4<k<24$ &(2,6,3,k)& $5<k<15$ & (2,7,3,k)&$6<k<12$ \\ \hline
 (2,8,3,k)&$7<k<11$ &(2,9,3,k) & $ 8<k<10$ &(4,2,3,k) &$3<k<6$ \\ \hline
 (2,k,3,7)& $7<k<22$ & (2,k,3,8) & $8<k<13$ & (k,2,3,3)& $3<k<6$ \\ \hline
 (3,4,2,k)& $3<k<18$ & (3,5,2,k) & $4<k<12$ & (3,6,2,k) & $5<k<10$ \\ \hline
 (3,7,2,k) & $6<k<9$ & (3,8,2,k) & $7<k<9$ & (4,3,2,k) &$ 3<k<8$ \\ \hline
 (3,k,2,7)& $7<k<15$ & (k,3,2,3) &$3<k<9$ & (k,3,2,4) & $4<k<6$ \\ \hline
 (k,4,2,3)& $4<k<8$ & (k,5,2,3) &$5<k<8$ &(k,6,2,3) & $6<k<8$ \\ \hline
 (2,4,4,k)& $3<k<12$ & (2,4,5,k) & $4<k<8$ &(2,5,4,k)& $4<k<8$ \\ \hline
 (2,6,4,k)& $5<k<7$ &(2,k,4,5)& $5<k<11$ &(4,4,2,k) & $3<k<6$ \\ \hline
 (4,5,2,k)& $4<k<6$ &(5,4,2,4)&(2,5,5,5)&(3,3,3,k)&$2<k<6$ \\ \hline
 (3,4,3,4)&(4,3,3,3)&&&& \\ 
    \hline
  \end{tabular}
    \caption{Sporadic sequence of type VIII singularity.}
  \label{VIIIfin}
\end{center}
\end{table}

\subsection{Type IX}
The singularity is $f=z_0^a+z_1^bz_3+z_2^cz_3+z_1z_3^d+z_2^pz_3^q$,  with constraint
${p(d-1)\over bd-1}+{qb(d-1)\over c(bd-1)}=1$. To have an isolated singularity at the origin, we 
need $a,b,d\geq 2$, $c\geq 1$. The simple reduction condition is that 
\begin{equation}
{b-1\over d-1}\in \mathbb{Z},~~~~{d-1\over b-1}\in \mathbb{Z},~~~~~{ c(b-1)\over b (d-1)}\in \mathbb{Z}
\end{equation}

The inequality we want to solve is
\begin{equation}
{1\over a}+ {d-1\over bd-1}+{b(d-1)\over c(bd-1)}+{b-1\over bd-1}>1,
\end{equation}
The infinite sequences of solutions (before imposing the constraint) are listed in table. \ref{IXinf}, and the sporadic 
solutions are listed in table. \ref{IXfin}.

\begin{table}[h]
\begin{center}
  \begin{tabular}{|c|c|c|c|c| }
    \hline
 (r,s,1,t)&(2,2,r,s)&(2,r,2,s)&(r,2,2,s)&(3,2,s,2)\\ \hline 
 (3,s,2,2)&(4,s,2,2)&(2,3,3,s)&(2,3,,4,s)&(2,3,5,s) \\ \hline
 (2,3,6,s)&(2,3,s,3)& (3,2,3,s)&(3,2,4,s)&(3,2,5,s) \\ \hline
 (3,2,6,s)&(2,4,3,s)&(2,5,3,s)&(2,6,3,s)&(4,2,3,s) \\ \hline
 (5,2,3,s)&(6,2,3,s)&(2,s,3,3)&(2,s,3,4)&(2,s,4,3) \\ \hline
 (3,3,2,s)&(3,4,2,s)&(3,5,2,s)&(3,6,2,s)&(4,3,2,s) \\ \hline
 (5,3,2,s)&(6,3,2,s)&(3,s,2,3)&(3,s,3,2)&(2,4,4,s) \\ \hline
 (4,2,4,s)&(4,4,2,s)&(3,3,3,s)&&\\ \hline
  \end{tabular}
  \caption{Infinite sequence of type IX singularity.}
  \label{IXinf}
\end{center}
\end{table}

\begin{table}[h]
\begin{center}
  \begin{tabular}{|c|c||c|c||c|c| }
    \hline
  (k,2,3,2)&$3<k<9$&(k,2,4,2)&$4<k<6$&(5,k,2,2)&$4<k<8$ \\ \hline 
  (k,3,2,2)&$3<k<10$&(k,4,2,2)&$4<k<7$&(2,3,7,k)&$6<k<15$ \\ \hline
  (2,3,8,k)& $7<k<9$ &(2,3,k,4)& $4<k<18$& (2,3,k,5)& $5<k<12$ \\ \hline
  (2,3,k,6)& $6<k<10$ &(2,3,k,7)& $7<k<9$ & (3,2,7,k)& $6<k<8$ \\ \hline
  (3,2,k,3)& $3<k<12$ &(3,2,k,4)& $4<k<9$ &(3,2,k,5)& $5<k<8$ \\ \hline
  (3,2,k,6)& $6<k<8$ & (2,7,3,k)& $6<k<19$ &(2,8,3,k)& $7<k<12$ \\ \hline
  (2,4,k,3)&$3<k<16$&(2,5,k,3)&$4<k<10$&(2,6,k,3)&$5<k<8$ \\ \hline
  (4,2,k,3)&$3<k<6$&(2,k,3,5)&$5<k<21$&(2,k,3,6)&$6<k<14$ \\ \hline
  (2,k,3,7)&$7<k<11$&(2,k,3,8)&$8<k<10$&(2,k,5,3)&$5<k<15$ \\ \hline
  (2,k,6,3)&$6<k<9$&(k,2,3,3)&$3<k<8$&(k,2,3,4)&$4<k<7$ \\ \hline
  (k,2,3,5)&$5<k<7$& (3,7,2,k)&$6<k<13$&(3,3,k,2)&$2<k<9$ \\ \hline
  (3,4,k,2)&$3<k<6$&(3,k,2,4)&$4<k<16$&(3,k,2,5)& $5<k<11$ \\ \hline
  (3,k,2,6)&$6<k<10$& (3,k,2,7)& $7<k<9$ & (3,k,4,2)&$4<k<8$ \\ \hline
  (k,3,2,3)&$3<k<8$& (k,3,2,4)& $4<k<8$ & (k,3,2,5)& $5<k<7$ \\ \hline
  (4,3,3,2)&(4,4,3,2)&(5,4,2,3)&(2,5,5,4)&(4,k,2,3)&$3<k<7$ \\ \hline
  (2,4,5,k)&$4<k<9$&(2,4,k,4)&$4<k<8$&(2,4,k,5)&$5<k<7$ \\ \hline
  (2,5,4,k)&$4<k<9$&(2,k,4,4)&$4<k<10$&(2,k,4,5)&$5<k<7$ \\ \hline
  (2,6,5,4)&(4,5,2,4)&(3,3,4,3)&(3,4,3,3)&& \\
    \hline
  \end{tabular}
  \caption{Sporadic sequence of type IX singularity.}
  \label{IXfin}
\end{center}
\end{table}

\subsection{Type X}
The singularity is $f=z_0^a+z_1^bz_2+z_2^c z_3+z_1 z_3^d$. We require $a\geq 2$ and $b,c,d\geq 2$ so that there is an isolated singularity at the origin. The singularity is reduced to previous type if
\begin{equation}
{b(d-1)+1\over d(c-1)+1}\in \mathbb{Z},~~\text{or}~~{c(b-1)+1\over b(d-1)+1}\in \mathbb{Z}~~\text{or}~~{d(c-1)+1\over c(b-1)+1}\in \mathbb{Z}.
\end{equation}
The inequality we would like to solve is: 
\begin{equation}
{1\over a}+ {d(c-1)+1\over bcd+1}+{b(d-1)+1\over bcd+1}+{c(b-1)+1\over bcd+1}>1.
\end{equation}
There is a symmetry in exchanging $(b,c,d)$ so we only label the solution with $b\leq c\leq d$. Notice that there are 
two inequivalent singularity labeled by $(b, c, d)$ and $(b, d, c)$ though (if two triples are related by cyclic permutation, then they define the same singularity).  
The infinite sequence of solutions are listed in table. \ref{Xinf}, and the sporadic sequence of solutions are listed in table. \ref{Xfin}.

\begin{table}[h]
\begin{center}
  \begin{tabular}{|c|c|c|c|c| }
    \hline
 (r,1,s,t)&(2,2,r,s)&(3,2,2,s)&(4,2,2,s)
&(s,2,2,2)\\ \hline 
(2,3,3,s)&(2,3,4,s)&(2,3,s,4)&(3,2,3,s)
 &(3,2,s,3) \\ \hline
  \end{tabular}
  \caption{Infinite sequence of type X singularity}
  \label{Xinf}
\end{center}
\end{table}
\begin{table}[h]
\begin{center}
  \begin{tabular}{|c|c||c|c||c|c| }
    \hline
  (5,2,2,k)&$4<k<11$&(6,2,2,k)&$5<k<7$&(k,2,2,3)&$3<k<13$ \\ \hline 
 (k,2,2,4)&$4<k<9$&(k,2,2,5)&$5<k<7$&(2,3,5,k)&$4<k<19$ \\ \hline
 (2,3,6,k)&$5<k<12$&(2,3,7,k)&$6<k<9$&(2,3,k,5)&$5<k<19$ \\ \hline
 (2,3,k,6)&$6<k<12$ &(2,3,k,7) &$7<k<9$&(3,2,4,k)&$3<k<13$ \\ \hline
 (3,2,5,k)&$4<k<8$ &(3,2,6,k)&$5<k<7$ &(3,2,k,4)& $4<k<13$ \\ \hline
 (3,2,k,5)&$5<k<8$ & (4,2,3,k)&$3<k<7$&(4,2,k,3)& $3<k<7$ \\ \hline
 (k,2,3,3)&$3<k<7$ &(2,4,4,k)&$3<k<11$&(2,4,5,k)& $4<k<7$ \\ \hline
 (2,4,6,5)& (4,2,4,4)&(3,3,3,k)& $2<k<6$&& \\
    \hline
  \end{tabular}
  \caption{Sporadic sequence of type X singularity}
  \label{Xfin}
\end{center}
\end{table}

\subsection{Type XI}
The singularity is $f=z_0^a+z_0z_1^b+z_1z_2^c+z_2z_3^d$.  We require $a\geq 2$, $b,c,d\geq 1$ 
so that there is an isolated singularity at the origin. 
The singularity is reduced to previous type if 
\begin{equation}
{b\over (a-1)}\in \mathbb{Z}~~\text{or}~~{ c(a-1)\over a(b-1)+1}\in \mathbb{Z},~~{d(a(b-1)+1)\over ab(c-1)+(a-1)}\in \mathbb{Z}. 
\end{equation}
The inequality we want to solve is 
\begin{equation}
{1\over a}+ {a-1\over ab}+{a(b-1)+1\over abc}+{ab(c-1)+(a-1)\over abcd}>1
\end{equation}
The infinite sequence of solutions are listed in table. \ref{XIinf}, and the sporadic sequence of solutions are listed in table. \ref{XIfin}.

\begin{table}[h]
\begin{center}
  \begin{tabular}{|c|c|c|c|c| }
    \hline
 (r,1,s,t)&(r,s,1,t)&(r,s,t,1)&(2,2,2,s)&(2,2,3,s) \\ \hline
 (2,2,s,2)&(2,2,s,3)&(2,2,s,4)&(2,r,2,s)&(3,2,2,s) \\ \hline
 (r,2,s,2)&(s,2,2,2)&(s,2,2,3)&(3,s,2,2)&(4,s,2,2) \\ \hline
 (s,3,2,2)&(2,3,s,3)&(3,2,s,3)&(2,s,3,3)&(2,s,3,4) \\ \hline
 (2,s,4,3)&(3,3,s,2)&(3,4,s,2)&(4,3,s,2)&(3,s,2,3) \\ \hline
 (3,s,3,2)&&&&\\ \hline
  \end{tabular}
  \caption{Infinite sequence of type XI singularity.}
  \label{XIinf}
\end{center}
\end{table}

\begin{table}[h]
\begin{center}
  \begin{tabular}{|c|c||c|c||c|c| }
    \hline
  (2,2,4,k)&$3<k<13$&(2,2,5,k)&$4<k<9$&(2,2,6,k)&$5<k<7$ \\ \hline 
 (2,2,k,5)&$5<k<12$&(2,2,k,6)&$6<k<8$&(4,2,2,k)&$3<k<11$ \\ \hline
 (5,2,2,k)&$4<k<7$&(k,2,2,4)&$4<k<11$&(k,2,2,5)&$5<k<7$ \\ \hline
 (5,k,2,2)&$4<k<12$&(6,k,2,2)&$5<k<8$&(k,4,2,2)&$4<k<13$ \\ \hline
 (k,5,2,2)&$5<k<9$&(2,3,3,k)&$2<k<13$ &(2,3,4,k)& $3<k<7$ \\ \hline
 (2,3,k,4)&$4<k<8$&(3,2,3,k)&$2<k<7$&(3,2,4,k)& $3<k<5$ \\ \hline
 (3,2,k,4)&$4<k<6$ &(2,4,3,k)& $3<k<9$ &(2,5,3,k)& $4<k<7$ \\ \hline
 (2,6,3,k)& $5<k<7$ & (2,4,k,3)& $3<k<13$ & (2,5,k,3)& $4<k<9$ \\ \hline
 (2,6,k,3)& $5<k<8$ & (4,2,3,4)&(5,2,5,3)&(4,2,k,3)& $ 3<k<10$ \\ \hline
 (2,k,3,5)& $5<k<11$ &(2,k,5,3)&$5<k<13$&(2,k,6,3)& $6<k<8$ \\ \hline
 (k,2,3,3)& $3<k<11$ &(k,2,4,3)& $4<k<7$&(3,3,2,k)& $2<k<11$ \\ \hline
 (3,4,2,k)& $3<k<7$ & (3,5,2,k)& $4<k<6$ &(3,5,k,2)& $4<k<13$ \\ \hline
 (3,6,k,2)& $5<k<8$ &(4,3,2,4)&(6,3,6,2)&(5,3,k,2)& $4<k<11$ \\ \hline
 (3,k,2,4)&$4<k<10$ &(3,k,4,2)& $4<k<14$ &(3,k,5,2)& $5<k<9$ \\ \hline
 (3,k,6,2)&$6<k<8$ & (k,3,2,3)&$3<k<7$ & (k,3,3,2)&$3<k<13$ \\ \hline
 (k,3,4,2)& $4<k<9$ &(k,3,5,2)& $5<k<7$ &(4,k,2,3)& $3<k<6$ \\ \hline
 (4,k,3,2) & $3<k<8$ & (k,4,3,2)& $ 4<k<7$ &(2,4,4,4)& (2,4,5,4) \\ \hline
 (2,k,4,4) &$4<k<7$ & (4,4,k,2) &$ 3<k<7$ & (4,5,4,2)& (3,3,3,3) \\ \hline
 (3,3,4,3)&(3,4,3,3)&&&\\
    \hline
  \end{tabular}
    \caption{Sporadic sequence of type XI singularity.}
  \label{XIfin}
\end{center}
\end{table}

\newpage
\subsection{Type XII}
Here $f=z_0^a+z_0z_1^b+z_0z_2^c+z_1z_3^d+z_1^pz_2^q$ with constraint ${p(a-1)\over ab}+{q(a-1)\over ac}=1$. 
The inequality we want to solve is 
\begin{equation}
{1\over a}+ {a-1\over ab}+{a-1\over ac}+{a(b-1)+1\over abd}>1
\end{equation}
The infinite sequence of solutions are listed in table. \ref{XIIinf}, while the sporadic solutions are listed in table. \ref{XIIfin}.
\begin{table}[h]
\begin{center}
  \begin{tabular}{|c|c|c|c|c| }
    \hline
 (r,1,s,t)&(r,s,1,t)&(r,s,t,1)&(2,2,2,s)&(2,2,s,2) \\ \hline
 (2,2,s,2)&(2,2,s,3)&(2,r,s,2)&(r,2,2,s)&(3,2,s,2) \\ \hline
 (2,s,2,2)&(2,s,2,3)&(2,s,2,4)&(s,2,3,2)&(s,2,4,2)\\ \hline
 (r,s,2,2)&(2,s,3,3)&(s,2,3,3)&(3,s,2,3)&(3,s,3,2) \\ \hline
 (3,s,4,2)&(s,3,2,3)&(s,3,2,4)&(s,3,3,2)&(4,s,3,2) \\ \hline
 (s,3,3,2)&(4,s,3,2)&(s,4,2,3)&& \\ \hline
  \end{tabular}
  \caption{Infinite sequence of type XII singularity}
  \label{XIIinf}
\end{center}
\end{table}

\begin{table}[h]
\begin{center}
  \begin{tabular}{|c|c||c|c||c|c| }
    \hline
  (2,2,3,k)&$2<k<9$&(2,2,4,k)&$3<k<6$&(2,2,k,4)&$4<k<8$ \\ \hline 
  (2,3,2,k)&$2<k<10$&(2,4,2,k)&$3<k<7$&(2,5,2,k)&$4<k<6$ \\ \hline
  (4,2,k,2)&$3<k<12$&(5,2,k,2)&$4<k<8$&(6,2,k,2)&$5<k<7$ \\ \hline
  (2,k,2,5)&$5<k<8$&(k,2,5,2)&$5<k<11$&(2,3,3,k)&$2<k<5$ \\ \hline
  (2,3,k,3)&$3<k<9$&(3,2,3,k)&$2<k<6$&(3,2,k,3)&$3<k<6$ \\ \hline
  (2,4,k,3)&$3<k<6$&(2,k,3,3)&$4<k<8$&(k,2,3,4)&$4<k<7$ \\ \hline
  (3,3,2,k)&$2<k<7$&(3,4,2,k)&$3<k<5$&(3,3,k,2)&$2<k<12$ \\ \hline
  (3,4,k,2)&$3<k<8$&(3,5,k,2)&$4<k<7$&(4,3,2,k)&$3<k<6$ \\ \hline
  (5,3,2,k)&$4<k<6$&(4,3,k,2)&$3<k<6$&(3,k,2,4)&$3<k<6$ \\ \hline
  (3,k,5,2)&$5<k<10$& (k,3,2,5)& $5<k<7$&(k,3,4,2)&$4<k<7$ \\ \hline
  (4,k,2,3)& $3<k<12$& (5,k,2,3)&$4<k<8$&(6,k,2,3)&$5<k<7$ \\ \hline
  (5,k,3,2)& $4<k<12$ & (6,k,3,2)&$5<k<8$& (k,5,2,3)& $5<k<9$ \\ \hline
  (4,4,2,4)&(4,4,4,2)&(4,5,4,2)&(3,3,3,3) \\
    \hline
  \end{tabular}
    \caption{Sporadic sequence of type XII singularity}
  \label{XIIfin}
\end{center}
\end{table}

\newpage
\subsection{Type XIII}
Here $f=z_0^a+z_0z_1^b+z_1z_2^c+z_1z_3^d+z_2^pz_3^q$ with constraint ${p(a(b-1)+1)\over abc}+{q(a(b-1)+1)\over abd}=1$. The inequality we want to solve is 
\begin{equation}
{1\over a}+ {a-1\over ab}+{a-1\over ac}+{a(b-1)+1\over abd}>1.
\end{equation}
The infinite sequence of solutions are listed in table. \ref{XIIIinf}, while the finite sequence of solutions are listed in table. \ref{XIIIfin}.
\begin{table}[h]
\begin{center}
  \begin{tabular}{|c|c|c|c|c| }
    \hline
 (r,1,s,t)&(r,s,1,t)&(r,s,t,1)&(2,2,2,s)&(2,2,3,s) \\ \hline
 (2,r,2,s)&(3,2,2,s)&(s,2,2,2)&(r,s,2,2)&(2,s,3,3) \\ \hline
 (2,s,3,4)&(2,s,3,5)&(2,s,3,6)&(3,s,2,3)&(3,s,2,4) \\ \hline
 (3,s,2,5)&(3,s,2,6)&(4,s,2,3)&(5,s,2,3)&(6,s,2,3) \\ \hline
 (2,s,4,4)&(4,s,2,4)&(3,s,3,3)&& \\ \hline
  \end{tabular}
  \caption{Infinite sequence of type XIII singularity.}
  \label{XIIIinf}
\end{center}
\end{table}
\begin{table}[h]
\begin{center}
  \begin{tabular}{|c|c||c|c||c|c| }
    \hline
  (2,2,4,k)&$3<k<12$&(2,2,5,k)&$4<k<8$&(4,2,2,k)& $3<k<10$ \\ \hline
  (5,2,2,k)&$4<k<8$ &(k,2,2,3)&$3<k<11$ &(k,2,2,4)& $4<k<7$ \\ \hline
  (2,3,3,k)&$2<k<15$ &(2,3,4,k)& $3<k<7$ &(3,2,3,k)& $2<k<6$ \\ \hline
  (2,4,3,k)&$3<k<11$ &(2,5,3,k)& $4<k<9$ & (2,6,3,k)&$5<k<9$ \\ \hline
  (2,7,3,k)& $6<k<8$ &(4,2,3,3)&(2,5,4,5)&(3,3,2,k)& $2<k<14$ \\ \hline
 (3,4,2,k)&$ 3<k<10$ & (3,5,2,k)&$4<k<9$ & (3,6,2,k)& $5<k<8$ \\ \hline
 (3,7,2,k)& $6<k<8$ & (4,3,2,k)& $3<k<6$ & (3,k,2,7)& $7<k<10$ \\ \hline
 (k,3,2,3)& $3<k<9$&(k,3,2,4)&$ 4<k<6$ &(k,4,2,3)& $4<k<8$ \\ \hline
 (k,5,2,3)&$5<k<8$ &(2,4,4,k)&$3<k<6$ &(4,4,2,k)& $3<k<6$ \\ \hline
 (3,3,3,k)& $2<k<5$ &&& \\ \hline
  \end{tabular}
    \caption{Sporadic sequence of type XIII singularity.}
  \label{XIIIfin}
\end{center}
\end{table}

\subsection{Type XIV}
Here $f=z_0^a+z_0z_1^b+z_0z_2^c+z_0z_3^d+z_1^pz_2^q+z_2^rz_3^s$ with constraint ${p(a-1)\over ab}+{q(a-1)\over ac}=1={r(a-1)\over ac}+{s(a-1)\over ad}$. The 
inequality we want to solve is 
\begin{equation}
{1\over a}+ {a-1\over ab}+{a-1\over ac}+{a-1\over ad}>1.
\end{equation}
The infinite sequence of solutions are listed in table. \ref{XIVinf}, while the sporadic ones are listed in table. \ref{XIVfin}.
\begin{table}[h]
\begin{center}
  \begin{tabular}{|c|c|c|c|c| }
    \hline
 (r,1,s,t)&(r,s,1,t)&(r,s,t,1)&(2,2,2,s)&(r,2,2,s) \\ \hline
 (s,2,3,3)&(s,2,3,4)&(s,2,3,5)&& \\ \hline
  \end{tabular}
  \caption{Infinite sequence of type XIV singularity.}
  \label{XIVinf}
\end{center}
\end{table}
\begin{table}[h]
\begin{center}
  \begin{tabular}{|c|c||c|c||c|c| }
    \hline
  (3,2,3,k)&$2<k<6$&(4,2,3,k)&$3<k<6$&(5,2,3,k)& $4<k<6$ \\ \hline
  \end{tabular}
    \caption{Sporadic sequence of type XIV singularity.}
  \label{XIVfin}
\end{center}
\end{table}

\subsection{Type XV}
Here $f=z_0^az_1+z_0z_1^b+z_0z_2^c+z_2z_3^d+z_1^pz_2^q$ with the constraint ${p(a-1)\over ab-1}+{qb(a-1)\over c(ab-1)}=1$. The inequality we want to solve is 
\begin{equation}
{b-1\over ab-1}+{a-1\over ab-1}+{b(a-1)\over c(ab-1)}+{c(ab-1)-b(a-1)\over cd(ab-1)}>1.
\end{equation}
The infinite sequence of solutions are listed in table. \ref{XVinf}, and the sporadic sequence of solutions are listed in table. \ref{XVfin}.
\begin{table}[h]
\begin{center}
  \begin{tabular}{|c|c|c|c|c| }
    \hline
(r,s,1,t)&(r,s,t,1)&(2,2,2,s)&(2,2,s,2)&(2,2,s,3) \\ \hline
(2,r,s,2)&(r,2,2,s)&(2,s,2,2)&(2,s,2,3)&(3,s,2,2) \\ \hline
(s,3,2,2)&(s,3,2,2)&(s,4,2,2)&(s,2,3,3)&(s,2,3,4) \\ \hline
(s,2,4,3)&(3,3,s,2)&(s,3,2,3)&(s,3,3,2)& \\ \hline
  \end{tabular}
  \caption{Infinite sequence of type XV singularity.}
  \label{XVinf}
\end{center}
\end{table}
There are following sporadic solutions
\begin{table}[h]
\begin{center}
  \begin{tabular}{|c|c||c|c||c|c| }
    \hline
  (2,2,3,k)&$2<k<7$&(2,2,4,k)&$3<k<5$&(2,2,k,4)& $4<k<6$ \\ \hline
  (2,3,2,k)&$2<k<7$&(2,4,2,k)&$3<k<5$&(2,k,2,4)&$4<k<6$ \\ \hline
  (4,k,2,2)&$3<k<10$&(5,k,2,2)&$4<k<7$&(k,5,2,2)&$5<k<9$ \\ \hline
  (2,3,k,3)&$3<k<6$&(3,2,3,k)&$2<k<6$&(3,2,k,3)&$3<k<8$ \\ \hline
  (4,2,k,3)&$3<k<6$&(5,2,k,3)&$4<k<6$&(2,k,3,3)&$3<k<6$ \\ \hline
  (3,3,2,k)&$2<k<5$&(3,4,k,2)&$3<k<8$&(4,3,k,2)&$3<k<9$ \\ \hline
  (5,3,k,2)&$4<k<6$&(3,k,3,2)&$3<k<9$&(3,k,4,2)&$4<k<6$ \\ \hline
  (k,3,4,2)&$4<k<9$&(2,3,3,3)&(4,2,3,4)&(4,3,2,4)&(3,4,2,3) \\ \hline
  (4,4,3,2)&(5,4,3,2)&&&& \\ \hline
  \end{tabular}
  \caption{Sporadic sequence of type XV singularity.}
  \label{XVfin}
\end{center}
\end{table}

\subsection{Type XVI}
Here $f=z_0^az_1+z_0z_1^b+z_0z_2^c+z_0z_3^d+z_1^pz_2^q+z_2^rz_3^s$ with the constraint ${p(a-1)\over ab-1}+{qb(a-1)\over c(ab-1)}=1={r(a-1)\over ac}+{s(a-1)\over ad}$. The 
inequality we want to solve is 
\begin{equation}
{b-1\over ab-1}+{a-1\over ab-1}+{b(a-1)\over c(ab-1)}+{b(a-1)\over d(ab-1)}>1;
\end{equation}
The infinite sequence of solutions are listed in table. \ref{XVIinf}, and the finite sequence of solutions are listed in table. \ref{XVIfin}.
\begin{table}[h]
\begin{center}
  \begin{tabular}{|c|c|c|c|c| }
    \hline
(r,s,1,t)&(r,s,t,1)&(2,2,2,s)&(r,2,2,s)&(2,s,2,2) \\ \hline
(r,s,2,2)&(s,2,3,3)&(s,2,3,4)&(s,2,3,5)&(s,3,2,3)\\ \hline
(s,3,2,4)&(s,3,2,5)&(s,4,2,3)&(s,5,2,3)& \\ \hline
  \end{tabular}
  \caption{Infinite sequence of type XVI singularity.}
  \label{XVIinf}
\end{center}
\end{table}

\begin{table}[h]
\begin{center}
  \begin{tabular}{|c|c||c|c||c|c| }
    \hline
  (2,2,3,k)&$2<k<6$&(2,3,2,k)&$2<k<6$&(2,k,2,3)&$3<k<6$ \\ \hline
  (3,2,3,k)&$3<k<6$&(4,2,3,k)&$3<k<6$&(5,2,3,k)&$4<k<6$ \\ \hline
  (3,3,2,k)&$2<k<6$&(4,3,2,k)&$3<k<6$&(5,3,2,k)&$4<k<6$ \\ \hline
  (3,k,2,3)&$3<k<6$&(4,k,2,3)&$3<k<6$&(5,k,2,3)&$4<k<6$ \\ \hline
  \end{tabular}
    \caption{Sporadic sequence of type XVI singularity.}
  \label{XVIfin}
\end{center}
\end{table}

\subsection{Type XVII}
Here $f=z_0^az_1+z_0z_1^b+z_1z_2^c+z_0z_3^d+z_1^pz_2^q+z_0^rz_2^s$ with the constraint 
${p(a-1)\over ab-1}+{qb(a-1)\over d(ab-1)}=1={r(b-1)\over ab-1}+{sa(b-1)\over c(ab-1)}$. The inequality we want to solve is 
\begin{equation}
{b-1\over ab-1}+{a-1\over ab-1}+{a(b-1)\over c(ab-1)}+{b(a-1)\over d(ab-1)}>1;
\end{equation}
The infinite sequence of solutions are listed in table. \ref{XVIIinf}, and the sporadic solutions are listed in table. \ref{XVIIfin}.
\begin{table}[h]
\begin{center}
  \begin{tabular}{|c|c|c|c|c| }
    \hline
(r,s,1,t)&(r,s,t,1)&(2,2,2,s)&(2,2,s,2)&(2,r,2,s)\\ \hline
(r,2,s,2)&(2,s,3,2)&(2,s,4,2)&(s,2,2,2)&(s,2,2,3) \\ \hline
(s,2,2,4)&(r,s,2,2)&(2,s,3,3)&(s,2,3,3)&(3,s,2,3) \\ \hline
(3,s,2,4)&(3,s,3,2)&(s,3,2,3)&(s,3,3,2)&(s,3,4,2) \\ \hline
(4,s,2,3)&(s,4,3,2)&&& \\ \hline
  \end{tabular}
  \caption{Infinite sequence of type XVII singularity.}
  \label{XVIIinf}
\end{center}
\end{table}

\begin{table}[h]
\begin{center}
  \begin{tabular}{|c|c||c|c||c|c| }
    \hline
  (2,3,k,2)&$2<k<8$&(2,4,k,2)&$3<k<6$&(2,5,k,2)&$4<k<6$ \\ \hline
  (3,2,2,k)&$2<k<8$&(4,2,2,k)&$3<k<6$&(2,3,3,k)&$2<k<5$ \\ \hline
  (3,3,2,k)&$2<k<6$&(3,4,2,k)&$3<k<6$&(3,3,k,2)&$2<k<6$ \\ \hline
  (3,4,k,2)&$3<k<5$&(4,3,k,2)&$3<k<6$&(5,k,2,3)&$4<k<9$ \\ \hline
  (4,k,3,2)&$3<k<10$&(5,k,3,2)&$4<k<7$&(k,4,2,3)&$4<k<10$ \\ \hline
  (k,5,2,3)&$5<k<7$&(k,5,3,2)&$5<k<9$&(3,2,3,3)&(3,2,4,3) \\ \hline
  (4,3,2,4)&&&&& \\ \hline
  \end{tabular}
   \caption{Sporadic sequence of type XVII singularity.}
  \label{XVIIfin}
\end{center}
\end{table}

\subsection{Type XVIII}
The singularity has the form $f=z_0^a z_2+ z_0 z_1^b+z_1z_2^c+z_1z_3^d+z_2^pz_3^q$ with the constraint 
${p(a(b-1)+1)\over abc+1}+{qc[a(b-1)+1]\over d(abc+1)}=1$. The inequality we want to solve is
\begin{equation}
{b(c-1)+1\over abc+1}+{c(a-1)+1\over abc+1}+{a(b-1)+1\over c(abc+1)}+{c(a(b-1)+1)\over d(abc+1)}>1;
\end{equation}
The infinite sequence of solutions are listed in table. \ref{XVIIIinf}, while the sporadic sequence of solutions are listed in table. \ref{XVIIIfin}. 
\begin{table}[h]
\begin{center}
  \begin{tabular}{|c|c|c|c|c| }
    \hline
(1,r,s,t)&(r,1,s,t)&(r,s,1,t)&(r,s,t,1)&(2,2,2,s) \\ \hline
(2,2,s,2)&(2,2,s,3)&(2,r,s,2)&(3,2,s,2)&(2,s,2,2) \\ \hline
(2,s,2,3)&(2,s,2,4)&(s,2,2,2)&(r,s,2,2)&(2,s,3,3) \\ \hline
(3,s,2,3)&(3,s,3,2)&(3,s,4,2)&(4,s,3,2)& \\ \hline
  \end{tabular}
  \caption{Infinite sequence of type XVIII singularity.}
  \label{XVIIIinf}
\end{center}
\end{table}
\begin{table}[h]
\begin{center}
  \begin{tabular}{|c|c||c|c||c|c| }
    \hline
  (2,2,3,k)&$2<k<9$&(2,2,4,k)&$3<k<6$&(2,2,k,4)&$4<k<8$ \\ \hline
  (2,3,2,k)&$2<k<10$ &(2,4,2,k)&$3<k<7$&(2,5,2,k)&$4<k<6$ \\ \hline
  (3,2,2,k)&$2<k<8$&(4,2,2,k)&$3<k<5$&(4,2,k,2)&$3<k<8$ \\ \hline
  (2,k,2,5)&$5<k<8$&(k,2,2,3)&$3<k<8$&(k,2,3,2)&$3<k<9$ \\ \hline
  (k,2,4,2)&$4<k<6$&(2,3,3,k)&$2<k<5$&(2,3,k,3)&$3<k<9$ \\ \hline
  (2,4,k,3)&$3<k<6$&(2,k,4,3)&$4<k<8$&(3,3,2,k)&$2<k<5$ \\ \hline
  (3,3,k,2)&$2<k<10$&(3,4,k,2)&$3<k<7$&(3,5,k,2)&$4<k<6$ \\ \hline
  (3,k,5,2)&$5<k<8$&(k,3,2,3)&$3<k<6$&(k,3,3,2)&$3<k<7$ \\ \hline
  (4,k,2,3)&$3<k<6$&(5,k,3,2)&$4<k<6$&(3,2,3,3)&(3,2,4,3) \\ \hline
  (2,4,3,4)&(4,3,4,2)&(5,4,3,2)&&\\ \hline
  \end{tabular}
   \caption{Sporadic sequence of type XVIII singularity.}
  \label{XVIIIfin}
\end{center}
\end{table}

\subsection{Type XIX}
The singularity is $f=z_0^az_2+z_0z_1^b+z_2^cz_1+z_2z_3^d$, and the inequality is 
\begin{equation}
({[b(d(c-1)+1)-1\over abcd-1},{[d(c(a-1)+1)-1\over abcd-1}
                ,{[a(b(d-1)+1)-1\over abcd-1},{[c(a(b-1)+1)-1\over abcd-1})>1.
\end{equation}
The infinite sequence of solutions are listed in table. \ref{XIXinf}, and the sporadic sequence of solutions 
are listed in table. \ref{XIXfin}.

\begin{table}[h]
\begin{center}
  \begin{tabular}{|c|c|c|}
    \hline
(1,r,s,t)&(2,r,s,2)&(2,2,s,3) \\ \hline
  \end{tabular}
  \caption{Infinite sequence of type XIX singularity.}
  \label{XIXinf}
\end{center}
\end{table}

\begin{table}[h]
\begin{center}
  \begin{tabular}{|c|c||c|c||c|c| }
    \hline
 (2,2,k,4)&$3<k<9$&(2,3,k,3)&$2<k<8$&(2,3,3,k)&$3<k<6$ \\ \hline
 (2,2,5,5)&(2,4,4,3)&&&& \\ \hline
  \end{tabular}
   \caption{Sporadic sequence of type XIX singularity.}
  \label{XIXfin}
\end{center}
\end{table}

\section{Beyond hypersurface singularity}
After classifying the hypersurface singularity, we would like to generalize the story to other type of singularities. We will discuss three other 
 constructions: a): Isolated complete intersection singularity (ICIS) with a $\mathbb{C}^*$ action. 
 b): If  there is a finite group action $G$ acting on ICIS and $G$ is also required to preserve the canonical three form, we can form the quotient and this seems to also 
define new theories; ; c): Instead of considering isolated singularity, we could consider non-isolated singularity.
Finally we conjecture that most general type of isolated singularity that would give us a $\mathcal{N}=2$ SCFT is rational graded Gorenstein singularity.

After describing the structure of singularity, there are three quantities which are of importance to us:
\begin{itemize}
\item $\mathcal{N}=2$ Geometry: The first object is the mini-versal deformation of the singularity, and we would like to determine the dimension of the base and 
the scaling dimension of the parameters parameterizing the base.
\item The second important quantity is the dimension of the middle homology group of Milnor fibration, and we would also like to determine the intersection form using distinguished 
basis associated with the vanishing cycle.
\item The third quantity is the number of co-dimension one singularities after the generic deformations. 
\end{itemize}
For Hypersurface singularity, the above three dimensions are the same and equal to the Milnor number $\mu$. For other more general singularities considered in this 
section, these three quantities are usually not the same. 
The story for ICIS is almost the same as the hypersurface case, and the quotient singularity 
has also been studied but less well understood, finally the non-isolated singularity has least understanding. 

We do not attempt here to provide full classification of all these constructions, we simply discuss the major features and provide simple examples, and 
we hope to classify them in the future work.

\subsection{Complete intersection}
The theory for isolated completed intersection singularity ( ICIS) can be found in \cite{arnol?d2012singularities, looijenga1984isolated}. 
The singularity is defined as the map $f:(\mathbb{C}^{n+3},0)\rightarrow (\mathbb{C}^n,0)$. Let's take the coordinates 
on $\mathbb{C}^{n+3}$ as $z_1,\ldots, z_{n+3}$, the singularity is defined as 
\begin{equation}
f_1(z_1,\ldots, z_{n+3})=f_2(z_1,\ldots, z_{n+3})=\ldots=f_{n}(z_1,\ldots,z_{n+3})=0
\end{equation}
We require $f_i$ to satisfy the following conditions:
\begin{itemize}
\item Each polynomial $f_i$ is quasi-homogeneous with degree $d_i$, and the weights of the coordinates $z_i$ are $w_i$. We assume that $d_i$ and $w_i$ are integers. 
\item The variety is required to be complete intersection, namely for the Jacobi matrix 
\begin{equation}
 {\partial f_a\over \partial z_i},~~a=1,\ldots, n,~~i=1,\ldots, n+3
\end{equation}
has rank $n$ everywhere except origin. 

\item There is an isolated singularity at the origin, namely, there is a unique solution $z_i=0$ for the equations
\begin{align}
& f_1=f_2=\ldots=f_n=0 \nonumber\\
& {\partial f_a\over \partial z_i}=0,~~a=1,\ldots, n,~~i=1,\ldots, n+3. 
\end{align}
\item The weights have to satisfy the condition $\sum w_i-\sum d_i>0$, and a derivation of this fact will be given later. 
\end{itemize}

The deformations are again related to the Jacobi module of the defining equations.
\begin{equation}
J(f)=C^n(z_0,z_1,\ldots z_{n+3})/({\partial f_a \over \partial z_j}).
\end{equation}
This vector space is finite dimensional if the singularity is isolated. 
Let's use f to denote the Column vector $(f_1,\ldots f_n)$, and $e_i$ to denote the basis of Jacobi module, then the mini-versal deformation of the singularity is 
\begin{equation}
F(z,\lambda)=f+\sum_{i=1}^\mu \lambda_i e_i,
\end{equation}
here again $\mu$ is the Milnor number which is equal to the dimension of the Jacobi module. 

We would like to find the scaling dimension for the coefficient before each deformation. As the $U(1)_R$ symmetry of $\mathcal{N}=2$ theory 
is proportional to the $\mathbb{C}^*$ action, we would like to find the proportional constant. To do that, we again require that the canonical three form has scaling dimension one. 
There is a canonical $(3,0)$ form which is defined as 
\begin{equation}
\Omega={dz_1\wedge dz_2\ldots \wedge d z_{n+3}\over df_1\wedge df_2\ldots \wedge df_n},
\end{equation}
The weights of $\Omega$ is $\sum_i w_i -\sum d_i$, and we require $\Omega$ to have scaling dimension one: 
\begin{equation}
(\sum_i w_i -\sum d_i)\delta=1,
\end{equation}
so $\delta={1\over (\sum_i w_i -\sum d_i)}$. For a deformation $\lambda_\alpha e_\alpha$, the scaling dimension would be 
\begin{equation}
[\lambda_\alpha]={d_\alpha-Q_{\alpha}\over (\sum_i w_i -\sum d_i)},
\label{ICIS}
\end{equation}
Here $d_\alpha$ is the degree of the polynomial $f_\alpha$ such that $\lambda_\alpha$ appears. 
We require $\Omega$ to have positive scaling dimension, so $\sum_i w_i-\sum_i d_i>0$ which is a simple generalization 
of the constraint on hypersurface singularity. 

Other physical quantities such as the central charges, low energy effective actions and the BPS quiver have the similar identifications as the geometric quantity, though the details
are often much more involved than the hypersurface case. We are not going to discuss any details here, and we only make two comments:  
The first comment is that the number of co-dimensional one singularities are not equal to the Milnor number, which is different from the hypersurface case. 
The number is $m=\mu+\mu^{'}$ with $\mu$ the Milnor number and $\mu^{'}$ is the Milnor number of ICIS defined by taking a polynomial out. This polynomial is chosen such that the 
remaining equations still define an isolated complete intersection singularity. The second comment is that the constraint $\sum w_i-\sum d_i$ is pretty strong, and we conjecture 
that maximal number of defining polynomials are two.

\textbf{Example}: Let's consider a complete intersection of two quadratics $I=(f_1,f_2)=(z_1^2+z_2^2+z_3^2+z_4^2, a_1z_1^2+a_2 z_2^2+a_3 z_3^2+a_4 z_4^2+a_5z_5^2)$, and $a_i$ is required to 
be distinct so that  there is an isolated singularity at the origin. The miniversal deformations of this singularity can be found using the software \textbf{Singular} \cite{DGPS}. The Milnor number is $9$, and 
the basis of the Jacobi module is 
\begin{equation}
 \left(\begin{array}{c}
0\\
z_5^2\end{array}\right),
 \left(\begin{array}{c}
0\\
z_5\end{array}\right),
 \left(\begin{array}{c}
0\\
z_4^2\end{array}\right),
 \left(\begin{array}{c}
0\\
z_4\end{array}\right),
 \left(\begin{array}{c}
0\\
z_3\end{array}\right),
 \left(\begin{array}{c}
0\\
z_2\end{array}\right),
 \left(\begin{array}{c}
0\\
z_1\end{array}\right),
 \left(\begin{array}{c}
0\\
1\end{array}\right),
 \left(\begin{array}{c}
1\\
0\end{array}\right).
\end{equation} 
So the miniversal deformations are 
\begin{align}
&(f_1+\lambda_1,f_2+\lambda_2 z_5^2+\lambda_3 z_5+\lambda_4 z_4^2+\lambda_5 z_4+\lambda_6 z_3+\lambda_7z_2+\lambda_8 z_1+\lambda_9). \nonumber\\
\end{align}
Using our formula for the scaling dimensions \ref{ICIS}, we find that there are five mass parameters, two operators with scaling dimension two and two operators with scaling dimension zero. 
There is a theory with the same type of spectrum: in class S theory, one can engineer a theory by putting 6d $A_1$ theory on a sphere with five regular punctures, and this theory 
has the same spectrum as the the above theory engineered using the ICIS. We compute the central charges of 
these two theories and they are equal, and this is a strong evidence that these two theories are the same. It would be interesting to find more evidence to check whether these two theories are the same or not.

\subsection{Quotient by finite group}
If the ICIS has finite group symmetries $G$  preserving the three form $\Omega$, we can consider new singularity formed by  quotient of 
this group $G$. For simplicity, we only consider hypersurface singularity, then the invariance under finite group $G$ implies that
\begin{equation}
f(T_g(z))=f(z).
\end{equation}
One can get a quotient space $X/G$ which also has an isolated singularity at the origin. Similarly one can
define the Milnor fibration of the quotient space, and we would like to determine the dimension $\hat{\mu}$ of the middle homology of the Milnor fibration. 
In the case of hypersurface, this number is equal to the dimension of Jacobi algebra of the singularity. In the quotient case, it is shown that 
the group action also acts on the homology, and the invariant subspace  is shown to be given by the following formula \cite{wall1980note}:
\begin{equation}
\hat{\mu}={1\over |G|}\sum_{g}(-1)^{d_g}\mu_g;
\end{equation}
Here $d_g$ is the codimension of the space in $\mathbb{C}^4$ which is fixed by the group element $g$, and $\mu_g$ is the dimension of the 
following vector space 
\begin{equation}
V_g/({\partial f\over z_i})_g;
\end{equation}
Here $V_g$ is the space invariant under the group element $g$ and $({\partial f\over z_i})_g$ is the Jacobi ideal which is 
invariant under $g$. This formula is the generalization of the formula of the hypersurface singularity in which the 
Milnor number is equal to the dimension of the Jacobi algebra. However, this is only part of middle homology groups of 
Milnor fibration, and we need to add more elements which is an analog of twisted sector in the study of orbifold. 
We do not have a complete story to discuss here, and in the following we give a simple example to illustrate 
the major point. 

\textbf{Example I}: Consider a case where there is a $\mathbb{Z}_2$ action acting on a single coordinates $z_3$, then there are two elements $(1,g)$, 
and the space fixed by $g$ is a three dimensional space specified by the coordinates $z_0,z_1,z_2$, so its codimension is one. Then the dimension of invariant part of homology 
group is given by $\hat{\mu}$: 
\begin{align}
& \hat{\mu}={1\over 2}(\mu-\mu_g); \nonumber\\
& \mu=\text{dim} {C(z_0,z_1,z_2,z_3)\over ({\partial f\over z_0}, {\partial f\over z_1},{\partial f\over z_2},{\partial f\over z_3})}; \nonumber\\
& \mu_g=\text{dim} {C(z_0,z_1,z_2)\over ({\partial f\over \partial z_0}|_{z_3=0}, {\partial f\over \partial z_1}|_{z_3=0},{\partial f\over \partial z_2}|_{z_3=0},{\partial f\over \partial z_3}|_{z_3=0})}.
\end{align}
This gives the number of vanishing three cycle, there are also hemi-cycles whose number can be determined by the fact that the dimension of middle homology of the Milnor vibration
is equal to the dimension of mini-versal deformation. 

We can give an explicit description for the deformation if 
there is only a $Z_2$ action acting on one of the coordinates $z_3$, then using 
the coordinates $\hat{z_3}=z_3^2$, the singularity is represented by a polynomial $\hat{f}$. The basis for the mini-versal deformation is 
given by the basis of the following Jacobi algebra
\begin{equation}
C(z_0,z_1, z_2, \hat{z}_3)/({\partial \hat{f}\over \partial z_0}, {\partial \hat{f}\over \partial z_1},{\partial \hat{f}\over \partial z_2},\hat{z_3}{\partial \hat{f}\over \partial \hat{z}_3})
\end{equation}
Let's use $\phi_i$ as the basis of this algebra, then the miniversal deformation is 
\begin{equation}
F(z_0,z_1,z_2, \hat{z}_3, \lambda)=f(z_0,z_1,z_2, \tilde{z}_3)+\sum_{i=1}^{\mu(\hat{f})} \lambda_i \phi_i.
\end{equation}
Once we find out the number $\mu(\hat{f})$, and we know the number of vanishing homology group, we can easily find out the number 
of hemispheres $\mu_{hemi}=\mu(\hat{f})-\hat{\mu}$. More general discussion on the quotient appearing in the Landau-Ginzburg context can be found in \cite{intriligator1990landau}.

\textbf{Example II}: Let's consider the following ADE singularity 
\begin{align}
&A_{2k-1}:~~~~f=z_0^2+z_1^2+z_2^2+z_3^{2k}, \nonumber\\
&D_{k+1}:~~~~~~f=z_0^2+z_1^2+z_2^{k}+z_2z_3^{2},  \nonumber\\
&E_6:~~~~~~~f=z_0^2+z_1^2+z_2^3+z_3^{4}.
\end{align}
There is a $Z_2$ acting on above singularity whose action is simply $Z_2:z_3\rightarrow -z_3$ (One can impose $z_0\rightarrow -z_0$ to ensure the invariance for the three form.). The new singularity and the miniversal deformations are 
\begin{align}
&B_k:~~\hat{f}=z_0^2+z_1^2+z_2^2+\hat{z}_3^{k},~~~\nonumber\\
&~~~~~~~F(z,\lambda)=\hat{f}+\lambda_1 \hat{z}_3^{k-1}+\ldots+ \lambda_{k-1} \hat{z_3}+\lambda_k, \nonumber\\
&C_k:~~~\hat{f}=z_0^2+z_1^2+z_2^{k}+z_2\hat{z}_3,  \nonumber\\
&~~~~~~~F(z,\lambda)=\hat{f}+\lambda_1 z_2^{k-1}+\ldots+ \lambda_{k-1} z_2+\lambda_k, \nonumber\\
&F_4:~~~\hat{f}=z_0^2+z_1^2+z_2^3+\hat{z}_3^{2}.  \nonumber\\
&~~~~~~~F(z,\lambda)=\hat{f}+\lambda_1 z_2+\lambda_2 \hat{z}_3+\lambda_3z_2\hat{z}_3+\lambda_4. \, \nonumber\\
\end{align}
The scaling dimension of the coefficients can be found from the original theory, i.e. the spectrum is a truncation of the original theory. 
We conjecture that those theories are the maximal AD points at the Coulomb branch of the corresponding pure $G$ gauge theory. The name is reflected 
in the fact that the intersection form of them have the same form as the corresponding Dynkin diagram. 
The remaining $G_2$ theory could be derived from taking a $Z_3$ quotient of $D_4$ singularity. We actually find several new rank one examples associated with $B_2, B_3, C_2, C_3, G_2$ 
Dynkin diagrams. Notice that $B_k$ and $C_k$ theory have the same spectrum from the SW curve, but their BPS quivers are different. We have seen before 
that for the hypersurface singularity, the spectrum of SW solution seem to complete characterize a theory, here we found theories which share the same $\mathcal{N}=2$ geometry, but with
different massive spectrum. It is interesting to further clarify what this means for the classification of $\mathcal{N}=2$ theory.

\subsection{General isolated singularity}
One might ask what is the most general type of isolated singularity that would give us a four dimensional $\mathcal{N}=2$ SCFT? As we described earlier, the necessary conditions are
\begin{itemize}
\item The singularity should have a $\mathbb{C}^*$ action to reflect the existence of $U(1)_R$ symmetry of the SCFT.
\item The SW geometry is described by the certain minimal deformation of the singularity, and the spectrum from the singularity should satisfy the pairing condition $[m]+[u]=2$. 
\item One need to have a SW differential defined on SW geometry to describe the mass of BPS particles, and we expect that this differential is also well defined on the singularity.
\end{itemize}
Let's now assume that the singularity is defined by an affine ring $R=\mathbb{C}[x_1, x_2, \ldots, x_n]/I$, and the first condition means that the ring is graded, i.e. there is a $\mathbb{C}^*$ action on the ring. 
The third condition means that the ring is Gorenstein \footnote{See \cite{eisenbud1995commutative} for the definition of Gorenstein ring.} so that there is a canonical form $\Omega$, and since there is a $\mathbb{C}^*$ action on the ring, $\Omega$ is graded and has weight 
$t$. We require that this grading to be positive 
\begin{equation}
t>0;
\end{equation}
mathematically this defines a rational singularity \cite{ishii1997introduction}. The integration of $\Omega$ over three cycles in the deformed geometry should give the mass of BPS particle, so we require the scaling dimension of it to be one. Using 
the above condition, we can find the proportional constant between the scaling dimension and the $\mathbb{C}^*$ charge:
\begin{equation}
t\delta=1\rightarrow \delta={1\over t}. 
\end{equation}
The deformation is now described by the Jacobi algebra, which is also graded. The scaling dimension of a coefficient before a basis vector $\phi_i$ in Jacobi algebra is 
\begin{equation}
[\lambda_i]={Q_i\over t};
\end{equation}
An extremely interesting property of this Jacobi algebra is that there is a perfect paring between the charges of the deformations \cite{wahl1987jacobian}
\begin{equation}
Q(m)+Q(u)=2t\rightarrow [m]+[u]=2.
\end{equation}
So a rational Gorenstein graded isolated three-fold singularity seems to define a $\mathcal{N}=2$ SCFT. 

For the hypersurface isolated singularity with a $\mathbb{C}^*$ action $f(\lambda^q_i z_i)=f(z_i)$, the rational condition is simply $\sum q_i>1$ as we discussed in section II. 
One must be careful that usually there is a stabilization of singularity $f(z_i)+w_1^2$, and the base of deformation space is not affected, namely the Jacobi algebra is 
the same, however, the property of rationally is not invariant under the stabilization. For example, a three dimensional singularity $z_0^2+z_1^3+z_2^7$ 
is not a two dimensional rational singularity, but its stabilization $z_0^2+z_1^3+z_2^7+z_3^2$ is a three dimensional rational singularity. 

The specially about the rational singularity is that the dual graph is homotopy to a point \cite{stepanov2008note}, and this might be the indication that we can get a local SCFT. 
From string theory point of view, this condition is equivalent to that the string theory on it is stable \cite{Giveon:1999zm}.
We are not aware of any other obstruction to the existence of SCFT. It would be interesting to further classify all possible $\mathcal{N}=2$ theory which can be engineered 
using three fold singularity. Among the interesting rational graded Gorenstein singularities are the toric singularities, quotient singularity $\mathbb{C}^3/G$ where $G$ is a finite subgroup of $SU(3)$, and 
we hope to come to study these singularities and its relation to $\mathcal{N}=2$ SCFT in the future.

\subsection{Non-isolated singularity and class ${\cal S}$ theory}
One could also consider the non-isolated singularity, and the interesting case is that there is a one-dimensional singular locus $\Sigma$. 
These type of singularities are less understood,  see \cite{arnolddynamical} for some preliminary discussions. 
We would like to rephrase the class ${\cal S}$ construction in the form of singularity theory. Consider the following singularity 
\begin{equation}
x^2+y^2+v^N(t^n+\sum_{i=1}^n c_i t^{n-1}+1)=0
\end{equation}
here $t$ is a $\mathbb{C}^*$ variable and  parameterizes a sphere with several punctures. 
If $f(t)=(t^n+\sum_{i=1}^n c_i t^{n-i}+1)\neq 0$, we get a $A_{N-1}$ singularity, so the singular locus is 
not a point but a one dimensional manifold. When $f(t)|_{t^*}=0$ the singularity behavior needs further 
study. This singularity has a $\mathbb{C}^*$ action if we assign charge $0$ to the $t$ variable, and 
the scaling dimension of $v$ would be simply $1$.  This singularity describes the following quiver:
\begin{equation}
N-\underbrace{SU(N)-SU(N)}_{n-1}-N
\end{equation}
The deformation of this singularity is described in \cite{Witten:1997sc}, and it takes the following form
\begin{equation}
x^2+y^2+\prod_{i=1}^N(v-m_{Li}) t^n+  \sum _{i=1}^{n-1}f_i(v) t^{n-i}+\prod_{i=1}^N(v-m_{Ri})=0
\end{equation}
with $f_v(i)=c_i v^N+ m_{i}v^{N-1}+ {u_{2i}} v^{N-2}+\ldots +u_{Ni}$.  The deformation theory of 
non-isolated singularity is quite complicated, and we hope that the above example could help us to 
understand better the non-isolated singularity.

\section{Conclusion}
Our philosophy of classifying 4d SCFT is similar to what is discussed in \cite{Argyres:2015ffa} in which they first define an isolated singularity with a good $\mathbb{C}^*$ action, and 
then constrain the possible deformations. In our case, we only specify the singularity, and the deformations are taken as the mini-versal deformations, namely the minimal deformations
such that all the other deformations are induced from it. It would be interesting to explore whether we can find new theories by allowing different deformation pattern. Alternatively, 
it might be the case that different deformation pattern actually gives us different singularities, and the deformation pattern is automatically encoded in 
the deformation theory of the singularities. 

For the hypersurface singularities, we have listed all possible solutions which can define a SCFT. It would be very interesting to completely classify other construction such 
as complete intersection, quotient by discrete group, etc. The non-isolated singularity is much less understood in mathematical literature, and it is definitely interesting 
to understand better this situation as it would teach us about the structure of class ${\cal S}$ theory. 

In this paper, our main approach is to use geometric singularity and its deformation to study 4d SCFT. Alternatively, it seems that we can define a 4d SCFT by using any 
two dimensional $(2,2)$ SCFT with central charge $\hat{c}<2$, and the deformation of the field theory is related to the $(c,c)$ chiral ring of 2d theory. It would be definitely interesting 
to try to classify 4d SCFT from this perspective.

The major point of this paper is to try to classify the singularities that lead to SCFT. There are many interesting physical questions such as structure of conformal manifold, 
RG flow, BPS spectrum, etc, and as we have shown here, the singularity theory plays an amazing role in understanding those physical quantities. The detailed 
study will be left for the other papers of this series. Other aspects such the calculation of superconformal index \cite{Gadde:2011uv,Buican:2015ina,Cordova:2015nma,Buican:2015tda,Song:2015wta} and associated chiral algebra \cite{Beem:2014rza} are also interesting to study from singularity theory point of view.

\section*{Acknowledgements}
We would like to thank 
Stephen S.T. Yau, Y.F. Wang and especially C. Vafa for helpful discussions. 
The work of S.T Yau is supported by  NSF grant  DMS-1159412, NSF grant PHY-
0937443, and NSF grant DMS-0804454.  
The work of DX is supported by Center for Mathematical Sciences and Applications at Harvard University, and in part by the Fundamental Laws Initiative of
the Center for the Fundamental Laws of Nature, Harvard University.

\bibliographystyle{utphys}

\bibliography{PLforRS}

\end{document}